\documentclass[10pt]{article}

\usepackage{amsmath}
\usepackage{amssymb}

\usepackage{graphicx}

\usepackage{cite}

\usepackage{color}

\usepackage[labelfont=bf,labelsep=period,justification=raggedright]{caption}

\makeatletter
\renewcommand{\@biblabel}[1]{\quad#1.}
\makeatother

\date{}

\begin{document}

\begin{flushleft}
{\Large
\textbf{The Population Genetic Signature of Polygenic Local Adaptation}
}
\\
Jeremy Berg$^{1,2,3,\ast}$, 
Graham Coop$^{1,2,3,\ast}$
\\
\bf{1} Graduate Group in Population Biology, University of California, Davis.
\\
\bf{2} Center for Population Biology, University of California, Davis.
\\
\bf{3} Department of Evolution and Ecology, University of California, Davis
\\
$\ast$ E-mail: jjberg@ucdavis.edu, gmcoop@ucdavis.edu
\end{flushleft}

\section*{Abstract}
Adaptation in response to selection on polygenic phenotypes may occur via subtle 
allele frequencies shifts at many loci.
Current population genomic techniques are not well posed to identify
such signals. 
In the past decade, detailed knowledge about the specific loci 
underlying polygenic traits has begun to emerge from genome-wide
association studies (GWAS). 
Here we combine this knowledge from GWAS with robust population
genetic modeling to identify traits that may have been influenced by local adaptation.
We exploit the fact that GWAS provide an estimate of the additive
effect size of many loci to estimate the mean additive genetic value for a given phenotype across 
many populations as simple weighted sums of allele frequencies.
We first describe a general model of neutral genetic value drift for an arbitrary number of populations
with an arbitrary relatedness structure. 
Based on this model we develop methods for detecting unusually strong
correlations between genetic values and specific environmental variables, as well as
a generalization of $Q_{ST}/F_{ST}$ comparisons to test for over-dispersion of genetic values among populations.
Finally we lay out a framework to identify the individual populations or groups of populations 
that contribute to the signal of overdispersion.
These tests have considerably greater power than their 
single locus equivalents due to the fact that they look for 
positive covariance between like effect alleles, and also significantly 
outperform methods that do not account for population structure.
We apply our tests to the Human Genome Diversity Panel (HGDP) dataset 
using GWAS data for height, skin pigmentation, type 2 diabetes, 
body mass index, and two inflammatory bowel disease datasets. 
This analysis uncovers a number of putative signals of local
adaptation, and we discuss the biological interpretation and caveats 
of these results.

\section*{Author Summary}

The process of adaptation is of fundamental importance in evolutionary biology. Within the last few decades, genotyping technologies and new statistical methods have given evolutionary biologists the ability to identify individual regions of the genome that are likely to have been important in this process. When adaptation occurs in traits that are underwritten by many genes, however, the genetic signals left behind are more diffuse, as no individual region of the genome will show strong signatures of selection. Identifying this signature therefore requires a detailed annotation of sites associated with a particular phenotype. Here we develop and implement a suite of statistical methods to integrate this sort of annotation from genome wide association studies with allele frequency data from many populations, providing a powerful way to identify the signal of adaptation in polygenic traits. We apply our methods to test for the impact of selection on human height, skin pigmentation, body mass index, type 2 diabetes risk, and inflammatory bowel disease risk. We find relatively strong signals for height and skin pigmentation, moderate signals for inflammatory bowel disease, and comparatively little evidence for body mass index and type 2 diabetes risk.

\section*{Introduction}

Population and quantitative genetics were in large part seeded by Fisher's
insight \cite{Fisher:1918wp} that the inheritance and evolution of quantitative characters 
could be explained by small contributions from many independent
Mendelian loci \cite{Provine:2001tf}. 
While still theoretically aligned \cite{Turelli:1990ub}, 
these two fields have often been divergent in empirical practice. 
Evolutionary quantitative geneticists have historically focused either on mapping the genetic basis of relatively simple traits \cite{Slate:2004fx}, or in the absence of any such knowledge, on understanding 
the evolutionary dynamics of phenotypes in response to selection over relatively short time-scales \cite{Kingsolver:2001di}.
Population geneticists, on the other hand, have usually focused on understanding the subtle signals left in genetic data by selection over longer time scales \cite{Hudson:1987wh,McDonald:1991hj,Begun:1992jd}, 
usually at the expense of a clear relationship between these patterns
of genetic diversity and evolution at the phenotypic level.





Recent advances in population
genetics have also allowed for the genome-wide identification of individual recent selective
events either by identifying unusually large allele frequency differences
among populations and environments or by detecting the effects of these events on linked diversity \cite{nielsen:2005bla}. 
Such approaches are nonetheless limited because they
rely on identifying individual loci that look unusual, and thus are
only capable of identifying selection on traits where an individual
allele has a large and/or sustained effect on fitness. When selection acts on a phenotype that is underwritten by a large number of loci,
the response at any given locus is expected to be modest, and the
signal instead manifests as a coordinated shift in allele frequency
across many loci, with the phenotype increasing alleles all on average
shifting in the same direction
\cite{Latta:1998tm,Latta:2003ed,LECORRE:2003vi,LeCorre:2012co,Kremer:2011kw}. Because
this signal is so weak at the level of the individual locus, it is
impossible to identify against the genome-wide background without a
very specific annotation of which sites are the target of selection on
a given trait \cite{Pritchard:2010wk}.
The advent of well-powered genome wide association studies with large sample sizes \cite{Risch} 
has allowed for just this sort of annotation, enabling the mapping of many small effect alleles associated
with phenotypic variation down to the scale of linkage disequilibrium in the population. 
The development and application of these methods in human populations has identified thousands of loci associated with a wide array of traits, largely confirming the polygenic view of phenotypic variation \cite{Visscher:2012je}.
%
%




Although the field of human medical genetics has been the largest and most rapid to puruse such approaches, evolutionary geneticists studying non-human model organisms have also carried out GWAS for a wide array of fitness-associated traits, and the development of further resources is ongoing \cite{Atwell2010,Fournier-Level2011,Mackay:2012fd}. 
In human populations, the cumulative contribution of these loci to the additive variance so far only explain a fraction of the narrow sense heritability for a given trait (usually less than 15\%), a phenomenon known as the missing heritability problem \cite{Manolio:2009jp,Bloom:2013bq}. Nonetheless, these GWAS hits represent a rich source of information about the loci underlying phenotypic variation.


Many investigators have begun to test whether the loci uncovered by these studies tend to be
enriched for signals of selection, in the hopes of learning more about how adaptation
has shaped phenotypic diversity and disease risk \cite{Myles:2008gg,Casto:2011di,paper:2013iua,Zhang:2013by}. The tests applied are generally still predicated on the idea of identifying individual loci that look unusual, such that a positive signal of selection is only observed if some subset of the GWAS loci have experienced strong enough selection to make them individually distinguishable from the genomic background. As noted above, it is unlikely that such a signature will exist, or at least be easy to detect, if adaptation is truly polygenic, and thus many selective events will not be identified by this approach.

Here we develop and implement a general method based on simple quantitative and population genetic principals, using allele frequency data at GWAS loci to test for a signal of selection on the phenotypes they underwrite while accounting for the hierarchical structure among populations induced by shared history and genetic drift. 
Our work is most closely related to the recent work of Turchin et al \cite{Turchin:2012dya}, Fraser \cite{Fraser:2013jja} and Corona et al \cite{Corona:2013iy}, 
who look for co-ordinated shifts in allele frequencies of GWAS alleles for particular traits.
Our approach constitutes an improvement over the methods implemented in these studies as 
it provides a high powered and theoretically grounded approach to investigate selection in 
an arbitrary number of populations with an arbitrary relatedness structure.


Using the set of GWAS effect size estimates and genome wide allele frequency data, 
we estimate the mean genetic value \cite{Fisher:GeneticalTheory,Falconer:QuantGen} for the trait of interest in 
a diverse array of human populations. These genetic values may in some cases be poor predictors of the actual phenotypes for reasons we address below and in the Discussion.
We therefore make no strong claims about their ability to predict present day observed phenotypes.
We instead focus on population genetic modeling of the joint distribution of genetic values,
which provides a robust way of investigating how selection may have impacted the underlying loci.

\begin{figure}[!ht]
	\includegraphics[width = \textwidth]{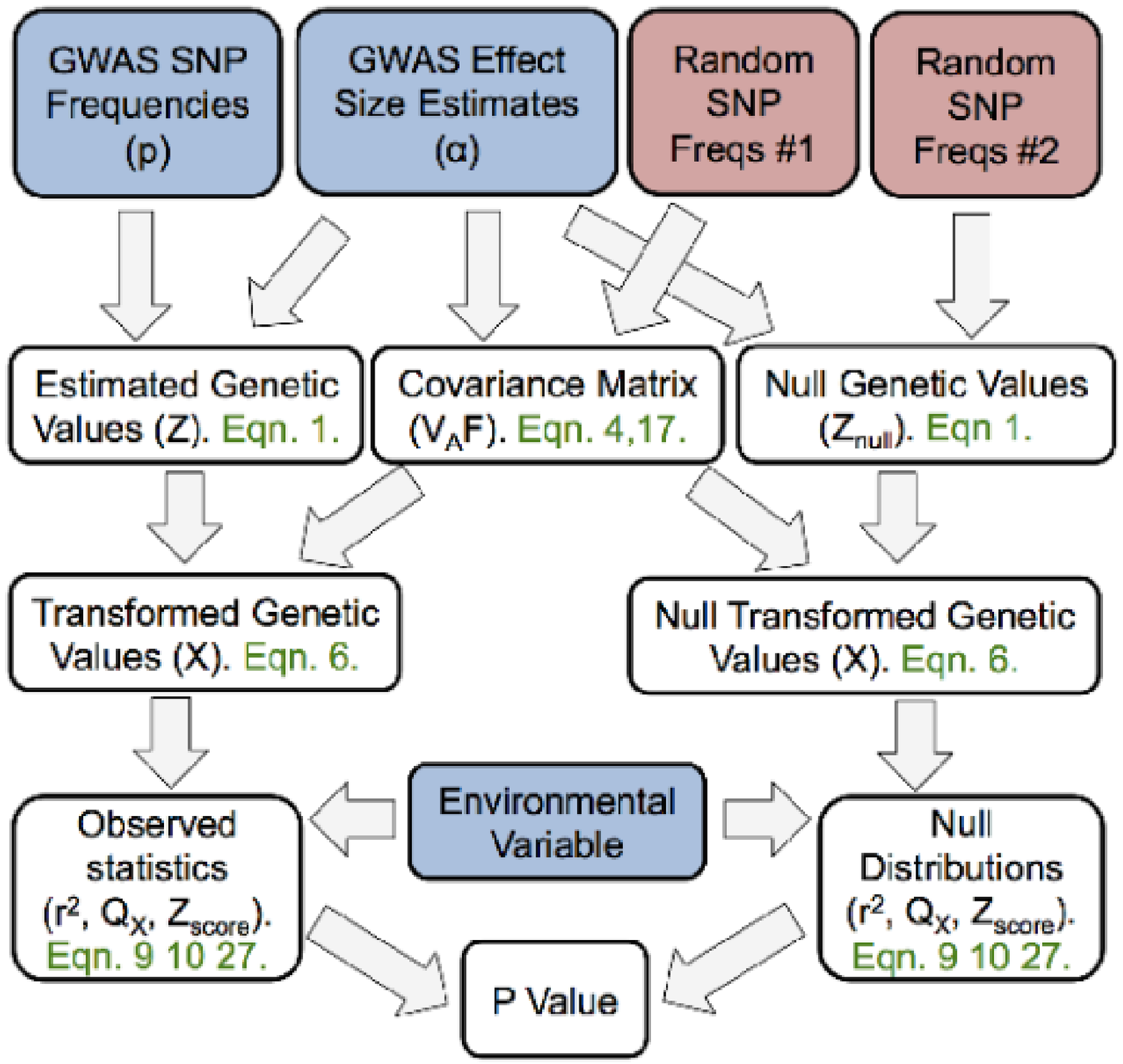}
	\caption{{\bf A schematic representation of the flow of our
          method.} The boxes colored blue are items provided by the
        investigator (GWAS SNP effect sizes, the frequency of the GWAS
        SNPs across populations, and a environmental variable). The boxes colored red make use of random SNPs sampled to match the
GWAS set as described in ``Choosing null SNPs'' in the methods
section. For each box featuring a calculated quantity a set of equation numbers are
provided for the relevant calculation. 
The Z score uses the untransformed genetic values,
rather than the transformed genetic values, but this relationship is not depicted in
the figure for the sake of readability. 
}
	\label{schematic}
\end{figure}

We develop a framework to describe how genetic values covary across populations based on a flexible 
model of genetic drift and population history. In Figure \ref{schematic} we show a
schematic diagram of our approach to aid the reader. Using this null model, 
we implement simple test statistics based on transformations 
of the genetic values that remove this covariance among populations.
We judge the significance of the departure from neutrality 
by comparing to a null distribution of test statistics constructed from well matched sets of control SNPs.
Specifically, we test for local adaptation by asking whether the transformed genetic values show excessive correlations with environmental or geographic variables. 
We also develop and implement a less powerful but more general test, which asks whether the genetic values are over-dispersed among populations compared to our null model of drift.
We show that this overdispersion test, which is closely related to $Q_{ST}$ \cite{Prout1993,Spitze1993} and a series of approaches from the population genetics literature \cite{Lewontin1973,MasatoshiNei:1975tt,Robertson:1975uo,Bonhomme2010,Gunther:2013iq},
gains considerable power to detect selection 
over single locus tests by looking for unexpected covariance among loci in the deviation they take from neutral expectations. 
Lastly, we develop an extension of our model that allows us to identify individual populations or groups of populations 
whose genetic values deviate from their neutral expectations given the
values observed for related populations, and thus have likely been
impacted by selection. 
While we develop these methods in the context of GWAS data, we also
relate them to recent methodological developments in the quantitative genetics of
observed phenotypes \cite{Ovaskainen:2011ku,Karhunen:2013eg},
highlighting the useful connection between these approaches. 

\section*{Results}
\subsection*{Estimating Genetic Values with GWAS Data}
\label{estimating-genetic-values}

Consider a trait of interest where $L$ loci (e.g. biallelic SNPs) have been identified from a genome-wide association study. We arbitrarily label the phenotype increasing allele $A_1$ and the alternate allele $A_2$ at each locus. These loci have additive effect size estimates $\alpha_1, \cdots \alpha_L$, where $\alpha_{\ell}$ is the average increase in an individual's phenotype from replacing an $A_2$ allele with an $A_1$ allele at locus $\ell$. We have allele frequency data for $M$ populations at our $L$ SNPs, and denote by $p_{m \ell}$  the observed sample frequency of allele $A_1$ at the $\ell^{th}$ locus in the $m^{th}$ population. From these, we estimate the mean genetic value in the $m^{th}$ population as
\begin{equation}
	Z_{m} = 2\sum_{\ell=1}^L\alpha_{\ell}p_{m\ell}
\end{equation}
and we take $\vec{Z}$ to be the vector containing the mean genetic values for all $M$ populations.

\subsection*{A Model of Genetic Value Drift}


We are chiefly interested in developing a framework for testing the hypothesis that the joint distribution of $\vec{Z}$ is driven by neutral processes alone, with rejection of this hypothesis implying a role for selection. We first describe a general model for the expected joint distribution of estimated genetic values ($\vec{Z}$) across populations under neutrality, accounting for genetic drift and shared population history.

A simple approximation to a model of genetic drift is that the current frequency of an allele in a population 
is normally distributed around some ancestral frequency ($\epsilon$). 
Under a Wright-Fisher model of genetic drift, the variance of this distribution is approximately 
$f \epsilon (1- \epsilon)$, where $f$ is a property of the population shared by all loci, 
reflecting the compounded effect of many generations of binomially sampling \cite{Nicholson:2002uk}. Note also that for small values, $f$ is approximately equal to the inbreeding coefficient of the present day population relative to the defined ancestral population, and thus has an interpretation as the correlation between two randomly chosen alleles relative to the ancestral population \cite{Nicholson:2002uk}. 


We can expand this framework to describe the joint distribution of allele frequencies across an arbitrary number of populations for an arbitrary demographic history by assuming that the vector of allele frequencies in $M$ populations follows a multivariate normal distribution
\begin{equation}
	\vec{p} \sim MVN\left(\epsilon\vec{1}, \epsilon\left(1-\epsilon\right)\mathbf{F} \right), \label{single_locus_mult_pop}
\end{equation}
where $\mathbf{F} $ is an $M$ by $M$ positive definite matrix describing the correlation structure of allele frequencies across populations relative to the mean/ancestral frequency. Note again that for small values it is also approximately the matrix of inbreeding coefficients (on the diagonal) and kinship coefficients (on the off-diagonals) describing relatedness among populations \cite{WEIR:2002gk,Bonhomme2010}. This flexible model was introduced, to our knowledge, by \cite{CavalliSforza:1964ea} (see\cite{Felsenstein:1982iw} for a review), and has subsequently been used as a computationally tractable model for population history inference \cite{Nicholson:2002uk,pickrell:2012fu}, and as a null model for signals of selection \cite{Coop2010,Bonhomme2010,Gunther:2013iq,Fariello:2013fv}.  So long as the multivariate normal assumption of drift holds reasonably well, this framework can summarize arbitrary population histories, including tree-like structures with substantial gene flow between populations \cite{pickrell:2012fu}, or even those which lack any coherent tree-like component, such as isolation by distance models \cite{Guillot:2012us,Bradburd:2013vf}.

Recall that our estimated genetic values $(\vec{Z})$ are merely a sum of sample allele frequencies weighted by effect size. If the underlying allele frequencies are well explained by the multivariate normal model described above, then the distribution of $\vec{Z}$ is a weighted sum of multivariate normals, such that this distribution is itself multivariate normal
\begin{equation}
	\vec{Z} \sim MVN\left( \mu\vec{1}, V_A \mathbf{F} \right )
\end{equation}
where $\mu = \frac{1}{L}\sum_{\ell=1}^L 2  \alpha_{\ell} \epsilon_{\ell}$ and $V_A = 4 \sum_{\ell=1}^L\alpha_{\ell}^2\epsilon_{\ell}(1-\epsilon_{\ell})$ are respectively the expected genetic value and additive genetic variance of the ancestral (global) population. The covariance matrix describing the distribution of $\vec{Z}$ therefore differs from that describing the distribution of frequencies at individual loci only by a scaling factor that can be interpreted as the contribution of the associated loci to the additive genetic variance present in a hypothetical population with allele frequencies equal to the grand mean of the sampled populations.

The assumption that the drift of allele frequencies around their shared mean is normally distributed \eqref{single_locus_mult_pop} may be problematic if there is substantial drift. However, even if that is the case, the estimated genetic values may still be assumed to follow a multivariate normal distribution by appealing to the central limit theorem, as each estimated genetic value is a sum over many loci. We show in the Results that this assumption often holds in practice.

It is useful here to note that the relationship between the model for drift at the individual locus level, and at the genetic value level, gives an insight into where most of the information and statistical power for our methods will come from. Each locus adds a contribution $2 \alpha_{\ell}(\vec{p}_{\ell } -\epsilon_\ell\vec{1})$ to the vector of deviations of the genetic values from the global mean. If the allele frequencies are unaffected by selection then the frequency deviation of an allele at locus $\ell$ in population $m$ $(p_{m,\ell } -\epsilon_\ell)$ will be uncorrelated in magnitude or sign with both the effect at locus $\ell$ $(\alpha_{\ell})$ and the allele frequency deviation taken by other unlinked loci. Thus the expected departure of a genetic value of a population from the mean is zero, and the noise around this should be well modeled by our multivariate normal model.

The tests described below will give positive results when these observations are violated. The effect of selection is to induce a non-independence of allele frequency deviation ($\vec{p}_{\ell } -\epsilon_\ell\vec{1}$) across loci, determined by the sign and magnitude of the effect sizes \cite{Latta:1998tm,Latta:2003ed,LECORRE:2003vi,Kremer:2011kw,LeCorre:2012co} and as we demonstrate below, all of our methods rely principally on identifying this non-independence. 
This observations has important considerations for the false positive
profile of our methods. Specifically, false positives will arise only if the GWAS ascertainment procedure induces a correlation between the estimated effect size of an allele ($\alpha_{\ell}$) and the deviation that this allele takes across populations $(\vec{p}_{\ell } -\epsilon_\ell\vec{1})$.  This should not be the case if the GWAS is performed in a single population which is well mixed compared to the populations considered in the test. False positives can occur when a GWAS is performed in a structured population and fails to account for the fact that the phenotype of interest is correlated with ancestry in this population. We address this case in greater depth in the Discussion.

These observation also allows us to exclude certain sources of statistical error as a cause of false positives. For example, simple error in the estimation of $\alpha_{\ell}$, or failing to include all loci affecting a trait cannot cause false positives, because this error has no systematic effect on $\vec{p}_{\ell } -\epsilon_\ell\vec{1}$ across loci. Similarly, if the trait of interest truly is neutral, variation in the true effects of an allele across populations or over time or space (which can arise from epistatic interactions among loci, or from gene by environment interactions) will not drive false positives, again because no systematic trends in population deviations will arise. This sort of heterogeneity can, however, reduce statistical power, as well as make straightforward interpretation of positive results difficult, points which we address further below.


\subsection*{Fitting the Model and Standardizing the Estimated Genetic Values}
As described above, we obtain the vector $\vec{Z}$ by summing allele frequencies across loci while weighting by effect size. We do not get to observe the ancestral genetic value of the sample $(\mu)$, 
so we assume that this is simply equal to the 
mean genetic value across populations $(\mu = \frac{1}{M}\sum_m Z_m)$. 
This assumption costs us a degree of freedom, and so we must work 
with a vector $\vec{Z^\prime}$, which is the vector of estimated genetic values for the first $M-1$ populations, centered at the mean of the $M$ (see Methods for details). Note that this procedure will be the norm for the rest of this paper, and thus we will always work with vectors of length $M-1$ that are obtained by subtracting the mean of the $M$ vector and dropping the last component.

To estimate the null covariance structure of the $M - 1$ populations we sample a large number K random unlinked SNPs.  In our procedure, the $K$ SNPs are sampled so as to match certain properties of the $L$ GWAS SNPs (the specific matching procedure is described in more depth below and in the Methods section).
Setting $\epsilon_k$ to be the mean sample allele frequency across populations at the $k^{th}$ SNP, we standardize the sample allele frequency in population $m$ as $(p_{mk}-\epsilon_{k})/\left(\epsilon_{k}\left(1-\epsilon_{k} \right)\right)$.  We then calculate the sample covariance matrix ($\mathbf{F}$) of these standardized frequencies, accounting for
the $M-1$ rank of the matrix (see Methods). 
We estimate the scaling factor of this matrix $\mathbf  F$ as
\begin{equation}
	V_A = 4\sum_{\ell=1}^L\alpha_{\ell}^2\epsilon_{\ell}(1-\epsilon_{\ell}).
\end{equation}




%


We now have an estimated genetic value for each population, and a simple null model describing their expected covariance due to shared population history. Under this multivariate normal framework, we can transform the vector of mean centered genetic values ($\vec{Z^\prime}$) so as to remove this covariance. First, we note that the Cholesky decomposition of the $\mathbf F$ matrix is
\begin{equation}
\mathbf F = \mathbf  C \mathbf{C^T}
\end{equation}
where $\mathbf  C$ is a lower triangular matrix, and $\mathbf {C^T}$ is its transpose. Informally, this can be thought of as taking the square root of $\mathbf  F$, and so $\mathbf{C}$ can loosely be thought of as analogous to the standard deviation matrix. 

 Using this matrix $\mathbf{C}$ we can transform our estimated genetic values as:
\begin{equation}
	\vec{X} = \frac{1}{\sqrt{V_A}}\mathbf C^{-1} \vec{Z^\prime}.
\end{equation}
If $\vec{Z^\prime}\sim \textrm{MVN}(\vec{0}, V_A \mathbf F)$ then $\vec{X} \sim \textrm{MVN}(0,\mathbb{I})$, where $\mathbb{I}$ is the identity matrix. Therefore, under the assumptions of our model, these standardized genetic values should be independent and identically distributed $\sim N(0,1)$ random variates \cite{Gunther:2013iq}.

It is worth spending a moment to consider what this transformation has done to the allele frequencies at the loci underlying the estimated genetic values. 
As our original genetic values are written as a weighted sum of allele frequencies, 
our transformed genetic values can be written as a weighted sum of transformed allele frequencies 
(which have passed through the same transform).
We can write 
\begin{equation}
	\vec{X} = \frac{1}{\sqrt{V_A}} \mathbf C^{-1} \vec{Z^\prime} = \frac{2}{\sqrt{V_A}} \sum_{\ell} \alpha_{\ell} \mathbf C^{-1} \vec{p}_{\ell} 
	\label{transformed-phenotypes}
\end{equation}
and so we can simply define the vector of transformed allele frequencies at locus $\ell$ to be 
\begin{equation}
	\vec{p^\prime}_{\ell} = \mathbf{C}^{-1} \vec{p}_{\ell}.
	\label{eq:transformed-freqs}
\end{equation}
This set of transformed frequencies exist within a set of transformed populations, 
which by definition have zero covariance with one another under the null, and are related by a star-like population tree with branches of equal length. 


As such, we can proceed with simple, straightforward and familiar statistical approaches to test for the impact of spatially varying selection on the estimated genetic values. Below we describe three simple methods for identifying the signature of polygenic adaptation, which arise naturally from this observation.

\subsection*{Environmental Correlations}
We first test if the genetic values are unusually correlated with an
environmental variable across populations compared to our null
model. A significant correlation is consistent with the hypothesis that the populations are
locally adapted, via the phenotype, to local conditions that are correlated with the environmental
variable. However, the link from correlation to causation must be supported by alternate forms of evidence,
and in the lack of such evidence, a positive result from our environmental correlation tests may be consistent with many explanations.

Assume we have a vector $\vec{Y}$, containing measurements of a specific environmental variable of interest in each of the $M$ populations.  We mean-center this vector and put it through a transform identical to that which we applied to the estimated genetic values in \eqref{transformed-phenotypes}. This gives us a vector $\vec{Y^\prime}$, which is in the same frame of reference as the transformed genetic values. 

There are many possible models to describe the relationship between a trait of interest 
and a particular environmental variable that may act as a selective agent.
We first consider a simple linear model, 
where we model the distribution of transformed genetic values ($\vec{X}$) 
as a linear effect of the transformed environmental variables ($\vec{Y^\prime}$)
\begin{equation}
	\vec{X} \sim \beta \vec{Y^\prime} + \vec{e} \label{environ_corr_eqn}
\end{equation}
where $\vec{e}$ under our null is a set of normal, independent and identically distributed random variates (i.e. residuals), 
and $\beta$ can simply be estimated as $\frac{Cov (
  \vec{X},\vec{Y'})}{Var ( Y ) }$.
We can also calculate the associated squared Pearson correlation coefficient ($r^2$) as a measure of the fraction of variance explained by our variable of choice, as well as the non-parametric Spearman's rank correlation $\rho\left(\vec{X},\vec{Y^\prime}\right)$, 
which is robust to outliers that can mislead the linear model. We note that we could equivalently pose this linear model as a mixed effects model,
with a random effect covariance matrix $V_A \mathbf{F}$. However, as
we know both $V_A$ and $\mathbf{F}$, we would not have to estimate any of the random effect parameters, reducing it to a fixed effect model
as in \eqref{environ_corr_eqn} \cite{RaoToutenburg}.

In the Methods (section ``The Linear Model at the Individual Locus Level'') we show that the linear environmental model applied to our transformed genetic values has a natural interpretation 
in terms of the underlying individual loci. Therefore, exploring the environmental correlations of estimated genetic values nicely summarizes information 
in a sensible way at the underlying loci identified by the GWAS.


In order to assess the significance of these measures, 
we implement an empirical null hypothesis testing framework, using $\beta$, $r^2$, and $\rho$ as test statistics. 
We sample many sets of $L$ SNPs randomly from the genome, 
again applying a matching procedure discussed below and in the Methods. 
With each set  of $L$ SNPs we construct a vector $\vec{Z}_{null}$, 
which represents a single draw from the genome-wide null distribution for a trait with the given ascertainment profile. 
We then perform an identical set of transformations and analyses on each $\vec{Z}_{null}$, 
thus obtaining an empirical genome-wide null distribution for all test statistics.

\subsection*{Excess Variance Test}
As an alternative to testing the hypothesis of an effect by a specific environmental variable, 
one might simply test whether the estimated genetic values exhibit more 
variance among populations than expected due to drift.
Here we develop a simple test of this hypothesis. 

As $\vec{X}$ is composed of $M-1$ independent, identically distributed standard normal random variables, a natural choice of test statistic is given by
\begin{equation} 
Q_X = \vec{X}^T \vec{X} = \frac{\vec{Z^\prime}^T\mathbf{F}^{-1}\vec{Z^\prime}}{V_A}.\label{eq:QxDef}
\end{equation}
This $Q_X$ statistic represents a standardized measure of the among population variance in estimated genetic values that is not explained by drift and shared history.
It is also worth noting that by comparing the rightmost form in \eqref{eq:QxDef} to the multivariate normal likelihood function,
we find that $Q_X$ is proportional to the negative log likelihood of the estimated genetic values under the neutral null model,
and is thus the natural measurement of the model's ability to explain their distribution.
Multivariate normal theory predicts that this statistic should follow a $\chi^2$ distribution 
with $M-1$ degrees of freedom under the null hypothesis.
Nonetheless, we use a similar approach to that described for the linear model, 
generating the empirical null distribution by resampling SNPs genome-wide.
As discussed below, we find that in practice the empirical null distribution tends to 
be very closely matched by the theoretically predicted $\chi^2_{M-1}$
distribution.

Values of this statistic that are in the upper tail 
correspond to an excess of variance among populations. This excess of
variance is
consistent with the differential action of natural selection on the phenotype
among populations  (e.g.\ due to local
adaptation). Values in the lower tail correspond a paucity of
variance, and thus potentially to widespread stabilizing selection, with many populations selected for the same optimum. 
In this paper we mainly concentrate on the upper tail of the distribution of
$Q_X$, e.g. for our power simulations, but note that either tail of
the distribution is informative about the action of selection on the phenotype.

\paragraph{The Relationship of $Q_X$ to Previous Tests} 

Our $Q_X$ statistic is closely related to $Q_{ST}$, the phenotypic analog of $F_{ST}$, which measures the fraction of the genetic variance that is among populations relative to the total genetic variance \cite{Wright1951,Spitze1993,Prout1993}. $Q_{ST}$ is typically estimated in traditional local adaptation studies via careful measurement of phenotypes from related individuals in multiple populations in a common garden setting. If the loci underlying the trait act in a purely additive manner and are experiencing only neutral genetic drift, then $\mathbb{E}[Q_{ST}] = \mathbb{E}[F_{ST}]$ \cite{Lande1992, Whitlock1999}.

If both quantities are well estimated, and we also assume that there is no hierarchical structure among the  populations, then $\frac{(M-1)Q_{ST}}{F_{ST}}$ is known to have a $\chi^2_{M-1}$ distribution under a wide range of models \cite{Rogers:1983wi,Whitlock2008,Whitlock:2009hf}. This statistic is thus a natural phenotypic extension of Lewontin and Krakauer's $F_{ST}$ based-test (LK test) \cite{Lewontin1973}.

To see the close correspondence between $Q_X$ and $Q_{ST}$, consider the case of a starlike population tree with branches of equal length (i.e. $f_{mm} = F_{ST}$ and $f_{m \neq n}=0$). Under this demographic model, we have
\begin{align}
	Q_X = \frac{\left(\vec{Z}-\mu\right)^T\mathbf{F}^{-1}\left(\vec{Z}-\mu\right)}{V_A} &= \frac{\left(Z_1 - \mu\right)^2}{V_A F_{ST}} + \dots + \frac{\left(Z_{M-1} - \mu\right)^2}{V_A F_{ST}} \notag \\
	& = \frac{\left(M-1\right)\text{Var}\left(\vec{Z}\right)}{V_A F_{ST}} \notag \\
	& = \frac{\left(M-1\right) \widehat{Q}_{ST}}{F_{ST}}
\end{align}
where $\widehat{Q}_{ST}$ is an estimated value for $Q_{ST}$ obtained from our estimated genetic values. This relationship between $Q_X$ and $Q_{ST}$ breaks down when some pairs of populations do not have zero covariance in allele frequencies under the null, in which case the $\chi^2$ distribution of the LK test also breaks down \cite{Robertson:1975uo,MasatoshiNei:1975tt}. Bonhomme and colleagues\cite{Bonhomme2010} recently proposed an extension to the LK test that accounts for a population tree, thereby recovering the $\chi^2$ distribution (see also \cite{Gunther:2013iq}, which relaxes the tree-like assumption), and our $Q_X$ statistic is a natural extension of this enhanced statistic to the problem of detecting coordinated selection at multiple loci. This test is also nearly identical to that developed by Ovaskainen and colleagues for application to direct phenotypic measurements \cite{Ovaskainen:2011ku}.



\paragraph{Writing $Q_X$ in Terms of Allele Frequencies}
Given that our estimated genetic values are simple linear sums of allele frequencies, 
it is natural to ask how $Q_X$ can be written in terms of these frequencies. Again, restricting ourselves to the case where $\mathbf{F}$ is diagonal,
(i.e. $f_{mm} = F_{ST}$ and $f_{m \neq n} = 0$), we can express $Q_X$ as
\begin{equation}
	Q_X = \frac{4}{V_A F_{ST}}\sum_{m=1}^{M-1} \sum_{\ell, \ell'} \alpha_{\ell}  \alpha_{ \ell'} (p_{m \ell} - \epsilon_{\ell}) (p_{m \ell'} - \epsilon_{\ell'}),
\end{equation}
which can be rewritten as
\begin{equation}
	Q_X = \frac{M - 1}{F_{ST}} \left(\frac{\sum_{\ell} \alpha_{\ell}^2 Var(\vec{p}_{\ell})}{\sum_{\ell} \alpha_{\ell}^2 \epsilon_{\ell} (1- \epsilon_{\ell} )} 
	+  \frac{\sum_{\ell \neq \ell '} \alpha_{\ell} \alpha_{\ell'} Cov(\vec{p}_{\ell},\vec{p}_{\ell '})}{\sum_{\ell}\alpha_{\ell}^2 \epsilon_{\ell} (1- \epsilon_{\ell} )} \right) \label{Q_X_as_FST}. \\
\end{equation}
The numerator of the first term inside the parentheses is the weighted sum of the variance among populations over all GWAS loci, scaled by the contribution of those loci to the additive genetic variance in the total population.
As such this first term is similar to $F_{ST}$ calculated for our GWAS loci, 
but instead of just averaging the among population and total variances equally across loci in the numerator and denominator, 
these quantities are weighted by the squared effect size at each locus. 
This weighting nicely captures the relative importance of different loci to the trait of interest.

The second term in  \eqref{Q_X_as_FST} is less familiar; the numerator is the weighted sum of the covariance of allele frequencies between all pairs of GWAS loci, and the denominator is again the contribution of those loci to the additive genetic variance in the total population.
This term is thus a measure of the correlation among loci in the deviation they take from the ancestral value, or the across population component of linkage disequilibrium. For a more in depth discussion of this relationship in the context of $Q_{ST}$, see \cite{Latta:1998tm,Latta:2003ed,LECORRE:2003vi,Kremer:2011kw,LeCorre:2012co}.

As noted above \eqref{eq:transformed-freqs}, when $\mathbf{F}$ is non-diagonal, our transformed genetic values can be written 
as a weighted sum of transformed allele frequencies. 
Consequently, we can obtain a similar expression to \eqref{Q_X_as_FST} when population structure exists,  
but now expressed in terms of the covariance of a set of allele frequencies
in transformed populations that have no covariance with each other under the null hypothesis. Specifically, when the covariance is non-diagonal we can write:
\begin{equation}
	Q_X = (M - 1) \frac{\sum_{\ell} \alpha_{\ell}^2 Var(\vec{p'}_{\ell})}{\sum_{\ell} \alpha_{\ell}^2 \epsilon_{\ell} (1- \epsilon_{\ell} )}  +
	(M - 1) \frac{\sum_{\ell \neq \ell '} \alpha_{\ell} \alpha_{\ell'} Cov(\vec{p'}_{\ell},\vec{p'}_{\ell '})}{\sum_{\ell} \alpha_{\ell}^2 \epsilon_{\ell} (1- \epsilon_{\ell} )}  \label{Q_X_two_comps}. \\
\end{equation}

We refer to the first term in this decomposition as the standardized $F_{ST}$-like component and the second term as the standardized LD-like component.
Under the neutral null hypothesis, the expectation of the second term is equal to zero, as
drifting loci are equally likely to covary in either direction.
With differential selection among populations, however, we expect loci underlying a trait not only to vary more than we would expect under a neutral model,
but also to covary in a consistent way across populations.
Models of local adaptation predict that it is this covariance among alleles that is primarily responsible for differentiation at the phenotypic level \cite{Latta:1998tm,Latta:2003ed,LECORRE:2003vi,LeCorre:2012co,Kremer:2011kw}, and we therefore expect the $Q_X$ statistic to offer considerably increased power as compared to measuring average $F_{ST}$ or identifying $F_{ST}$ outliers. We use simulations to demonstrate this fact below, and also demonstrate the perhaps surprising result that for a broad parameter range the standardized LD-like component exhibits almost no loss of power when used as a test statistic.

\subsection*{Identifying Outlier Populations}
\label{Section:Outlierpops}
Having detected a putative signal of selection for a given trait, one may wish to identify individual regions and populations which contribute to the signal. Here we rely on our multivariate normal model of relatedness among populations, along with well understood methods for generating conditional multivariate normal distributions, in order to investigate specific hypotheses about individual populations or groups of populations. Using standard results from multivariate normal theory, 
we can generate the expected joint conditional distribution of genetic values for an arbitrary set of populations 
given the observed genetic values in some other set of populations. 
These conditional distributions allow for a convenient way to ask whether the estimated genetic values 
observed in certain populations or groups of populations differ significantly from the values we would 
expect them to take under the neutral model given the values observed in related populations. 

Specifically, we exclude a population or set of populations,
and then calculate the expected mean and variance 
of genetic values in these excluded populations given the values 
observed in the remaining populations, and the covariance matrix relating them.
Using this conditional mean and variance, we calculate a Z-score to 
describe how well fit the estimated genetic values of the excluded populations
are by our model of drift, conditional on the values in the remaining populations.
In simple terms, the observation of an extreme Z-score for a particular population 
or group of populations may be seen as evidence that that group has experienced 
directional selection on the trait of interest (or a correlated one) that 
was not experienced by the related populations on which we condition the analyses.
The approach cannot uniquely determine the target of selection, however. 
For example, conditioning on populations that have themselves 
been influenced by directional selection may lead to large Z-scores for
the population being tested, even if that population has been evolving neutrally.
We refer the reader to the Methods section for a mathematical explication of these approaches.


\subsection*{Datasets}
We conducted power simulations and an empirical application of our methods based on the Human Genome Diversity Panel (HGDP) population genomic dataset \cite{Li:2008wj}, and a number of GWAS SNP sets. To ensure that we made the fullest possible use of the information in the HGDP data, we took advantage of a genome wide allele frequency dataset of $\sim$3 million SNPs imputed from the Phase II HapMap into the 52 populations of the HGDP. These SNPs were imputed as part of the HGDP phasing procedure in \cite{Pickrell:2009if}; see our Methods section for a recap of the details. We applied our method to test for signals of selection in six human GWAS datasets identifying SNPs associated with height, skin pigmentation, body mass index (BMI), type 2 diabetes (T2D), Crohn's Disease (CD) and Ulcerative Colitis (UC). 


\paragraph{Choosing null SNPs}
\label{Results:choosing-null-snps}
Various components of our procedure involve sampling random sets of
SNPs from across the genome. While we control for biases in our test
statistics introduced by population structure through our $\mathbf{F}$
matrix, we are also concerned that subtle ascertainment effects of the
GWAS process could lead to biased test statistics, even under neutral
conditions. We control for this possibility by sampling null SNPs so
as to match the joint distribution of certain properties of the
ascertained GWAS SNPs. Specifically, we chose our random SNPs to match
the GWAS SNPs in each study in terms of their minor allele frequency
(MAF) in the ascertainment population and the imputation status of the
allele in our population genomic dataset (i.e. whether the allele was
imputed or present in the original HGDP genotyping panel). In addition, 
we were concerned that GWAS SNPs might be preferentially found
 close to genes and in low recombination regions, the latter due to better tagging, 
and as such may be subject to a high rate of drift due to background selection, 
leading to higher levels of differentiation at these sites \cite{Charlesworth:1997wz}. 
Therefore, in addition to MAF and imputation status, we also matched our random SNPs to an estimate of the background selection environment experienced by the GWAS SNPs, as measured by B value \cite{McVicker:2009ic}, which is a function of both the density of functional sites and recombination rate calibrated to match the reduction in genetic diversity due to background selection. We detail the specifics of the binning scheme for matching the discretized distributions of GWAS and random SNPs in the Methods.

\subsection*{Power Simulations}

To assess the power of our methods in comparison to other possible approaches, we conducted a series of power simulations. There are two possible approaches to simulate the effect of selection on large scale allele frequency data of the type for which our methods are designed. The first is to simulate under some approximate model of the evolutionary history (e.g. full forward simulation under the Wright-Fisher model with selection). The second is to perturb real data in such a way that approximates the effect of selection. We choose to pursue the latter, both because it is more computationally tractable, and because it allows us to compare the power of our different approaches for populations with evolutionary histories of the same complexity as the real data we analyze. Each of our simulations will thus consist of sampling 1000 sets of SNPs matched to the height dataset (in much the same way we sample SNPs to construct the null distributions of our test statistics), and then adding slight shifts in frequency in various ways to mimic the effect of selection.

Below we first describe the set of alternative statistics to which we compare our methods. We then describe the manner in which we add perturbations to mimic selection, and lastly describe a number of variations on this theme which we pursued in order to better demonstrate how the power of our statistics changes as we vary parameters of the trait of interest, evolutionary process, or the ascertainment.

\paragraph{Statistics Tested}
For our first set of simulation experiments we compared two of our statistics, ($r^2$ and $Q_X$) 
against their naive counterparts, which are not adjusted for population structure (\textit{naive} $r^2$ and $Q_{ST}$). We also include the adjusted $F_{ST}$-like and LD-like components of $Q_X$ as their behavior over certain parameter ranges is particularly illuminating.
For $Q_{ST}$, $Q_X$, and it's components, we count a given simulation as producing a positive result if the statistic lies in the upper 5\% tail of the null distribution, whereas for the environmental correlation statistics ($r^2$ and \textit{naive} $r^2$) we use a two-tailed 5\% test.
We also compared our tests to a single locus enrichment test, where we tested for an enrichment in the number of SNPs that individually show a correlation with the environmental variable. We considered this test to produce a positive result if the number of
individual loci in the 5\% tail of the null distribution for individual locus $r^2$ was itself in the 5\% tail using a binomial test. We do not include our alternative linear model statistics $\beta$ and $\rho$ in these plots for the sake of figure legibility, but they generally had very similar power to that of $r^2$.
While slightly more powerful versions of the $r^2$ \textit{enrichment} test that better
account for sampling noise are available \cite{Coop2010}, note that
our tests could be extended similarly as well, so the comparison is
fair.

\begin{figure}[!ht]
	\includegraphics[height = \textheight]{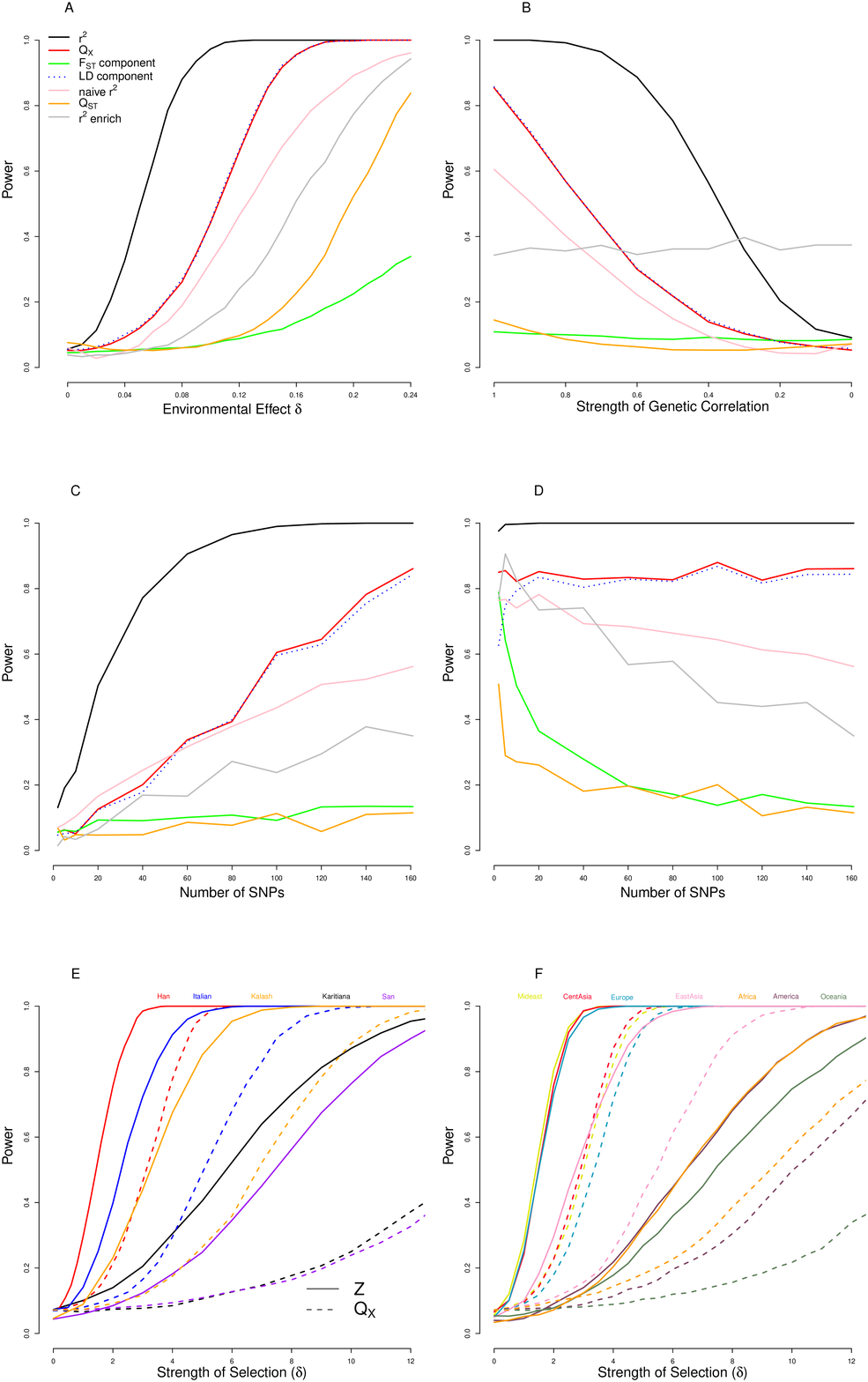}
	\caption{\textbf{Power of our statistics as compared to alternative approaches} (A) across a range of selection gradients ($\delta$) of latitude, and when we hold $\delta$ constant at 0.14 and (B) decrease $\phi$, the genetic correlation between the trait of interest and the selected trait, (C) vary the number of loci, and (D) vary the number of loci while holding the fraction of variance explained constant. Bottom panels show power of the Z-test and $Q_X$approaches to detect selection affecting (E) a single population, and (F) multiple populations in a given region. See main text for simulation details.}
	\label{power-plot}
\end{figure}

\paragraph{Simulating Selection}
We base our initial power simulations on empirical data altered to have an increasing effect of selection along a latitudinal gradient. In order to mimic the effect of selection, we generate a new set of allele frequencies ($p_{s,m\ell}$) by taking the original frequency ($p_{m \ell}$) and adding a small shift according to
\begin{align}
	p_{s,m\ell} = p_{m\ell} + p_{m\ell} \left(1-p_{m\ell}\right)\alpha_{\ell}\delta Y_m
	\label{add-selection}
\end{align}
where $\alpha_\ell$ is the effect size assigned assigned to locus $\ell$, and $Y_m$ is the mean centered absolute latitude of the population. We use 1000 simulations at $\delta=0$ to form null distribution for each of our test statistics, and from this established the $5\%$ significance level. We then increment $\delta$ and give the power of each statistic as the fraction of simulations whose test statistic falls beyond this cutoff. While this approach to simulating selection is obviously naive to the way selection actually operates, it captures many of the important effects on the loci underlying a given trait. Namely, loci will have greater shifts if they experience extreme environments, have large effects on the phenotype, or are at intermediate frequencies. Because we add these shifts to allele frequencies sampled from real, putatively neutral loci, the effect of drift on their joint distribution is already present, and thus does not need to be simulated. The results of these simulations are shown in Figure \ref{power-plot}A.

Our population structure adjusted statistics clearly outperform tests that do not account for structure, as well as the single locus outlier based test. Particularly noteworthy is the fact that the power of a test relying on $Q_X$ and that using only the LD-like component are essentially identical over the entire range of simulation, while the $F_{ST}$-like component achieves only about $20\%$ power by the point at which the former statistics have reached 100\%. This reinforces the observation from previous studies of $Q_{ST}$ that for polygenic traits, nearly all of the differentiation at the trait level arises as a consequence of across population covariance among the underlying loci, and not as a result of substantial differentiation at the loci themselves \cite{Kremer:2011kw}. While our environment-genetic value correlation tests considerably outperform $Q_{X}$, this is somewhat artificial as it assumes that we know the environmental variable responsible for our allele frequency shift. In reality, the power of the environmental variable test will depend on the investigator's ability to accurately identify the causal variable (or one closely correlated with it) in the particular system under study, and thus in some cases $Q_X$ may have have higher power in practice. Panels A and B from Figure \ref{power-plot} with SNPs matched to each of the other traits we investigate can be found in Figures \ref{skin-power-plot}-\ref{uc-power-plot}.

\paragraph{Pleiotropy and Correlated Selection}
We next considered the fact that many of the loci uncovered by GWAS are may be relatively pleiotropic, and thus may simultaneously respond to selection on multiple different traits. To explore how our methods perform in the presence of undetected pleiotropy, we consider the realization that from the perspective of allele frequency change there is only one effect that matters, and that is the effect on fitness. We therefore chose a simple and general approach to capture a flavor of this situation. We simulate the effect of selection as above \eqref{add-selection}, but give each locus an effect on fitness ($\alpha_\ell^\prime$) that may be only partially correlated with the observed effect sizes for the trait of interest (with the unaccounted for effect on fitness coming via pleiotropic relationships to any number of unaccounted for phenotypes). For simplicity we assume that $\alpha_{\ell}$ and $\alpha_{\ell}^\prime$ have a bivariate normal distribution around zero with equal variance and correlation parameter $\phi$. We then simulate $\alpha_{\ell}^\prime$ from its conditional distribution given $\alpha_{\ell}$ (i.e. $\alpha_{\ell}^\prime \mid \alpha_{\ell} \sim N(\phi \alpha_{\ell},(1-\phi^2) Var(\alpha) )$).
For each SNP $\ell$ in \eqref{add-selection} we replaced $\alpha_{\ell}$ by its effect $\alpha_{\ell}'$ on the unobserved phenotype, but then perform our tests using the $\alpha_\ell$ measured for the trait of interest. Here $\phi$ can be thought of as the genetic correlation between our phenotype and fitness 
if this simple multivariate form held true for all of the loci contributing to the trait.
The extremes of $\phi = 1$ and $\phi = 0$ respectively represent the
cases where selection acts only on the focal trait and that were all the underlying loci are affected by
selection, but not due to their relationship with the focal trait. These simulations can also informally 
be seen as modeling the case where the GWAS estimated
effect sizes are imperfectly correlated with the true effect sizes
that selection sees, for example due to measurement error in the GWAS.

In Figure \ref{power-plot}B we hold the value of $\delta$ constant at 0.14 
and vary the genetic correlation $\phi$ from one down to zero. 
Predictably, our GWAS genetic value based statistics lose 
power as the the focal trait becomes less correlated with fitness
but do retain reasonable power out to quite low genetic correlations
(e.g. our $r^2$ out performs the single locus metrics  until $\phi <0.3$). 
In contrast, counting the number of SNPs that are significantly
correlated with a given environmental variable remains equally
powerful across all genetic correlations. This is because the single locus environmental correlation tests 
treat each locus separately with no regards to whether there is
agreement across alleles with the same direction of effect size.
This may be a desirable property of the environmental outliers
enrichment approach, as it does not rely on a close relationship
between the effect sizes and the way that selection acts on the loci. On the other hand, this is also problematic, 
as such tests may often be detecting selection on only very weakly
pleiotropically related phenotypes. Our approaches, however, are more
suited to determining whether the genetic basis of a trait of
interest, or a reasonably correlated trait,   
has been affected by differentiating selection.


\paragraph{Ascertainment and Genetic Architecture}
We next investigated the relationship between statistical power,
the number of loci associated with the trait, and the amount of
variance explained by those loci. Our simulations were motivated by
the fact that the number of loci identified by a given GWAS, and the
fraction of variance explained by those loci, will depend on both the
design of the study (e.g. sample size) and the genetic architecture of the trait. To
illustrate the impact these factors have on the power of our methods,
we performed two experiments in which we again held $\delta$ constant at 0.14. 
In the first, for each of the 1000 sets
of 161 loci chosen above to mimic the height data ascertainment, we
randomly sampled $n$ loci, without regard to effect sizes, and recalculated the null distribution
and power for these reduced sets, allowing $n$ to range from 2 to
161. The results of these simulations are shown in Figure \ref{power-plot}C. This corresponds to imagining that fewer loci had been ascertained by the
initial GWAS, and estimating the power our methods would have with
this reduced set of loci. As we down sample our loci without regard to
effect sizes, the horizontal axis of Figure \ref{power-plot}C is proportional 
to the phenotypic variance explained, e.g. the simulations in which only 
80 loci are subsampled correspond to having a dataset which explains only 50\% of the 
variance explained in those for which all 161 were used.


The second experiment is nearly identical to the first, except that
before adding an effect of selection to the subsampled loci, we
linearly rescale the effect sizes such that $V_A$ is held constant at
the value calculated for the full set of 161 loci.  The results of these simulations are shown in Figure \ref{power-plot}D. These simulations
correspond to imagining that we have explained an equivalent amount of
phenotypic variance, but the number of loci over which this variation is
partitioned varies.

Our results (Figure \ref{power-plot}C and \ref{power-plot}D) demonstrate that
even if only a small number of loci associated with the
phenotype have been identified, our tests offer higher power than
single locus-based tests. Moreover, for statistics that appropriately
deal with both covariance among loci and among populations ($r^2$ and
$Q_X$), power is generally a constant function of variance explained
by the underlying loci, regardless of the number of loci over which 
it is partitioned. Notably, most the power of $Q_X$ comes from the LD-like
component, especially when the number of loci
is large. Statistics that rely on an average of single locus metrics (the $F_{ST}$-like component of
$Q_X$), and those that rely on outliers ($r^2$ \textit{enrichment})
all lose power as the the variance explained is partitioned over more
loci, as the effect of selection at each locus is weaker. Somewhat
surprisingly, the versions of our tests that fail to adequately control for
population structure (\textit{naive} $r^2$ and $Q_{ST}$) also lose
power as the phenotypic variance is spread among more loci. We believe
this reflects the fact that they are being systematically mislead by
LD among SNPs due to population structure, a problem which is
compounded as more loci are included in the test. Overall these
results suggest that accounting for population structure and using the
LD between like effect alleles is key to detecting selection on
polygenic phenotypes.

\paragraph{Localizing Signatures of Selection}
Lastly, we investigated the power of our conditional Z-scores to identify signals of selection that are specific to particular populations or geographic regions, and contrast this with the power of the global $Q_X$ statistic to detect the same signal. We again perform two experiments. In the first, we choose a single population whose allele frequencies to perturb, and leave all other populations unchanged. In other words, an effect of selection is mimicked according to \eqref{add-selection}, but with $Y_m$ set equal to one for a single population, and zero for all others. We then increment $\delta$ to see how power changes as the effect of selection becomes more pronounced. In Figure \ref{power-plot}E we display the results of these simulations for five populations chosen to capture the range of power profiles for the populations we consider in our empirical applications. In the last experiment, we chose a group of populations to which to apply the allele frequency shift, again consistent with \eqref{add-selection}, but now with $Y_m$ set equal to 1 for all populations in an entire region, and zero elsewhere. In Figure \ref{power-plot}F, we show the results of these simulations, with each of the seven geographic/genetic clusters identified by Rosenberg et al (2002) \cite{Rosenberg:2002ga}, chosen in turn as the affected region.

These simulations demonstrate that the conditional Z test can detect subtler frequency shifts than the global $Q_X$ test, provided one knows which population(s) to test \textit{a priori}. They also show how unusual frequency patterns indicative of selection are easier to detect in populations for which the dataset contains closely related populations that are unaffected (e.g. compare the Han and Italian to the San and Karitiana at the individual population level, or Europe, the Middle East and Central Asia to Africa, America, and Oceania at the regional level). Lastly, note that the horizontal axes in Figure \ref{power-plot}E and \ref{power-plot}F are equivalent in the sense that for a given value of $\delta$, alleles in (say) the Italian population have been shifted by the same amount in the Italian specific simulations in Figure \ref{power-plot}E as in the Europe-wide simulation in Figure \ref{power-plot}F, indicating that the HGDP dataset, power is similar in efforts to detect local, population specific events, as well as broader scale, regional level events.

\subsection*{Empirical Applications}
We estimated genetic values for each of six traits from the subset of GWAS SNPs that were present in the HGDP dataset, as described above. We discuss the analysis of each dataset in detail below, and address general points first. 
For each dataset, we constructed the covariance matrix from a sample of approximately $20,000$ appropriately matched SNPs, and the null distributions of our test statistics from a sample of $10,000$ sets of null genetic values, which were also constructed according to a similar matching procedure (as described in the Methods).


In an effort to be descriptive and unbiased in our decisions about which environmental variables to test, we tested each trait for an effect of the major climate variables considered by Hancock et al (2008) \cite{Hancock:2008cn} in their analysis of adaptation to climate at the level of individual SNPs. We followed their general procedure by running principal components (PC) analysis for both seasons on a matrix containing six major climate variables, as well as latitude and longitude (following Hancock et al's rationale that these two geographic variables may capture certain elements of the long term climatic environment experienced by human populations). The percent of the variance explained by these PCs and their weighting (eigenvectors) of the different environmental variables are given in Table \ref{tab:eigenvector-table}. We view these analyses largely as a descriptive data exploration enterprise across a relatively small number of phenotypes and distinct environmental variables, and do not impose a multiple testing penalty against our significance measures. A multiple
testing penalization or false discovery rate approach may be needed when testing a large number
phenotypes and/or environmental variables.

\begin{figure}[!ht]
	\includegraphics[width = \textwidth]{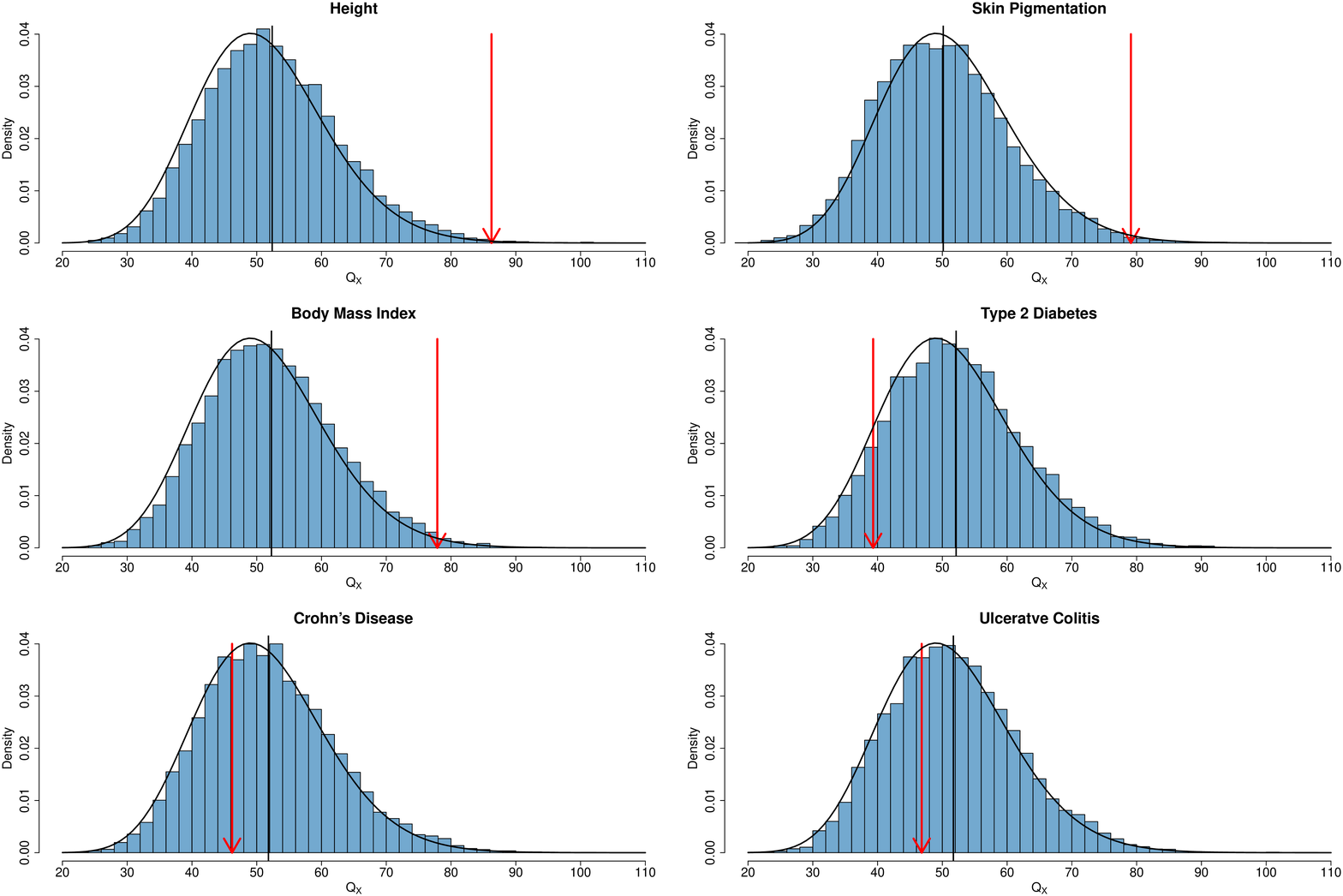}
	\caption{{\bf Histogram of the empirical null distribution of $Q_X$ for each trait, obtained from genome-wide resampling of well matched SNPs.} The mean of each distribution is marked with a vertical black bar and the observed value is marked by a red arrow. The expected $\chi^2_{M-1}$ density is shown as a black curve.}
	\label{QxHistograms}
\end{figure}

We also applied our $Q_X$ test to identify traits whose underlying loci showed consistent patterns of unusual differentiation across populations.  In Figure \ref{QxHistograms} 
we show for each GWAS set the observed value of $Q_X$  and its empirical null distribution calculated using SNPs matched to the GWAS loci as described above. We also plot the expected null distribution of the $Q_X$ statistic ($\sim \chi_{51}^2$). The expected null distribution closely matches the empirical distribution in all cases, 
suggesting that our multivariate normal framework provides a good null model for the data 
(although we will use the empirical null distribution to obtain measures of statistical significance).

\begin{figure}[!ht]
	\includegraphics[width = \textwidth]{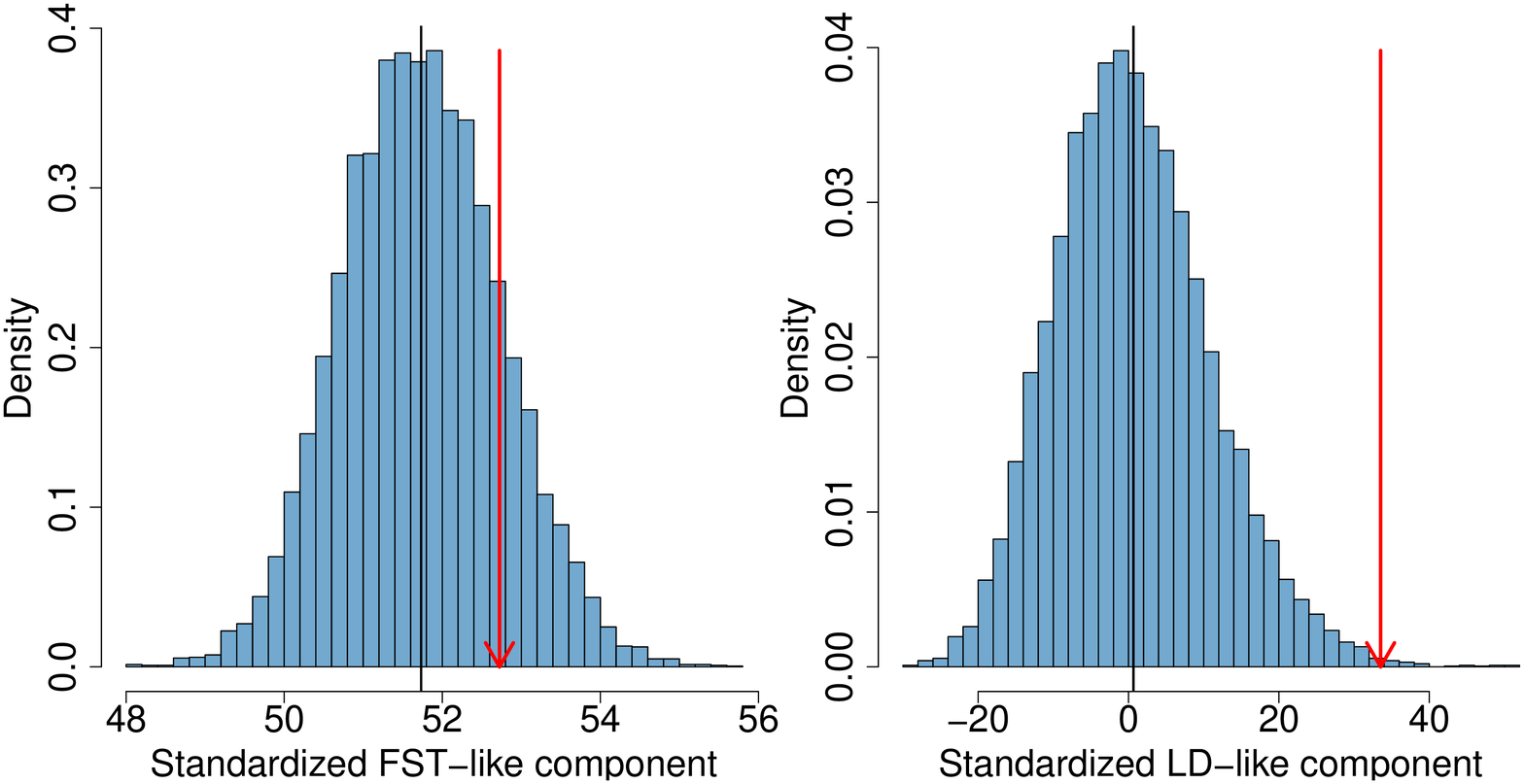}
	\caption{{\bf The two components of $Q_X$ for the height dataset, as described by the left and right terms in \eqref{Q_X_two_comps}}. The null distribution of each statistic is shown as a histogram. The mean value is shown as a black bar, and the observed value as a red arrow.}
	\label{QxComponents}
\end{figure}

For each GWAS SNP set we also separate our $Q_X$ statistic into its $F_{ST}$-like and LD-like terms, 
as described in \eqref {Q_X_two_comps}.
In Figure \ref{QxComponents} we plot the null distributions of these two components for the height dataset as histograms, 
with the observed value marked by red arrows (Figures \ref{Q_X_components_skin}-\ref{Q_X_components_UC} give these plots for the other five traits we examined). 
In accordance with the expectation from our power simulations, 
the signal of selection on height is driven entirely by covariance among loci in their deviations from neutrality, 
and not by the deviations themselves being unusually large.

\begin{figure}[!ht]
		\includegraphics[height = 0.9\textheight]{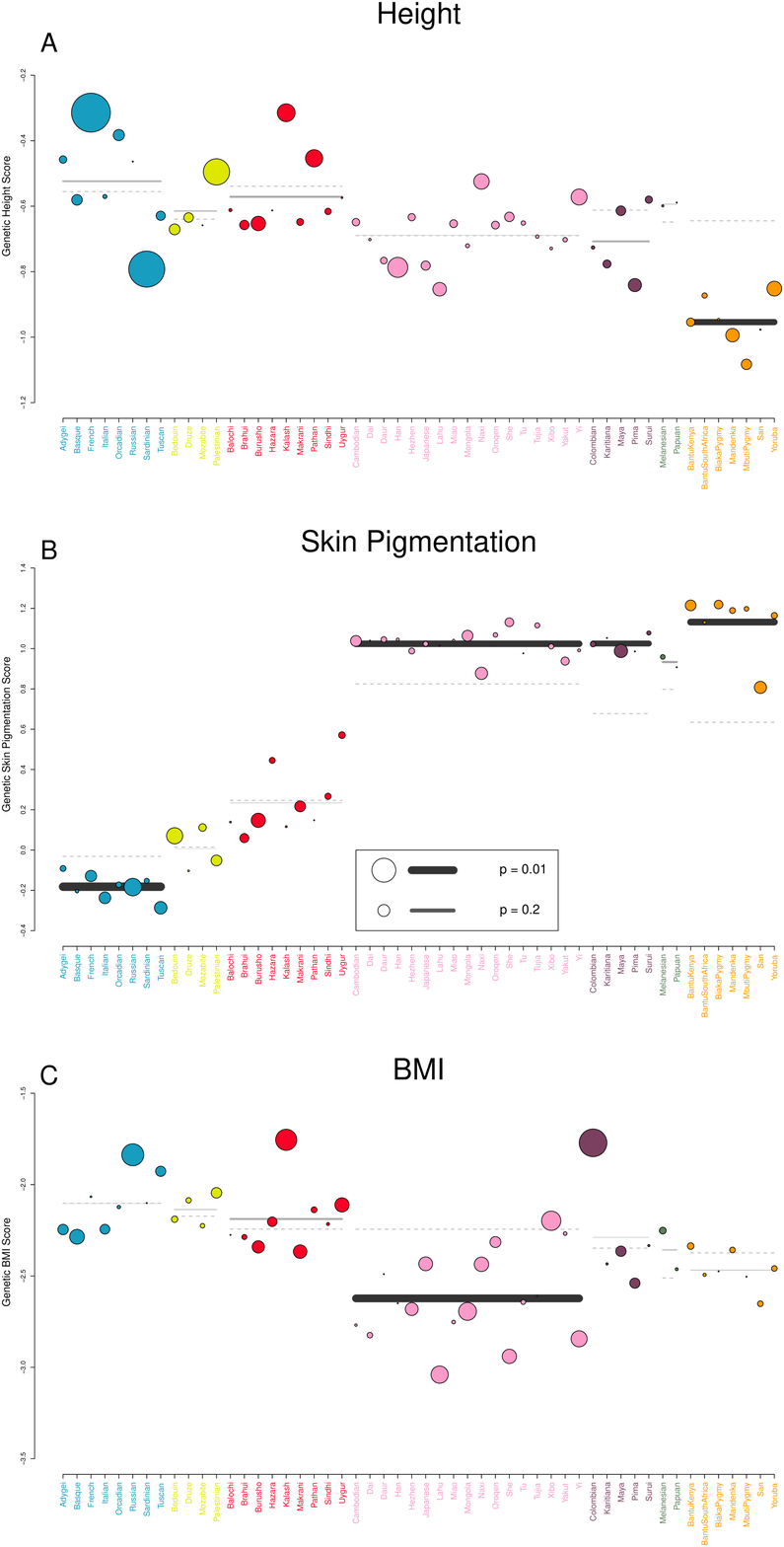}
		\caption{{\bf Visual representation of outlier analysis at the regional and individual population level for (A) Height, (B) Skin Pigmentation, and (C) Body Mass Index.} For each geographic region we plot the expectation of the regional average, given the observed values in the rest of the dataset as a grey dashed line. The true regional average is plotted as a solid bar, with darkness and thickness proportional to the regional Z score. For each population we plot the observed value as a colored circle, with circle size proportional to the population specific Z score. For example, in (A), one can see that estimated genetic height is systematically lower than expected across Africa. Similarly, estimated genetic height is significantly higher (lower) in the French (Sardinian) population than expected, given the values observed for all  other populations in the dataset.}\label{fig:height-skin-bmi-conditionals}
\end{figure}

\begin{figure}[!ht]
	\includegraphics[height = \textheight]{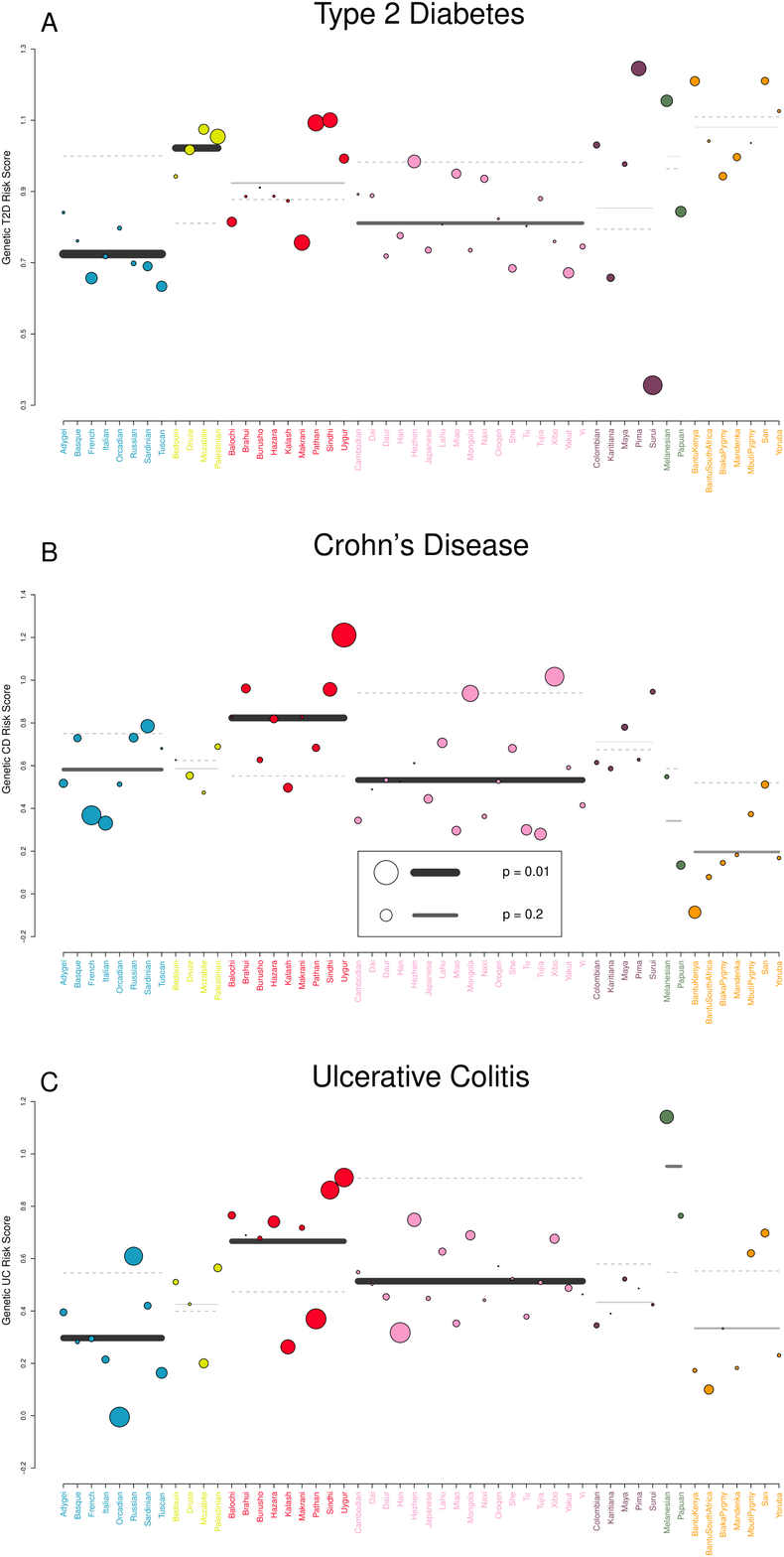}
	\caption{{\bf Visual representation of outlier analysis at the regional and individual population level for (A) Type 2 Diabetes, (B) Crohn's Disease, and (C) Ulcerative Colitis.} See Figure \ref{fig:height-skin-bmi-conditionals} for explanation. }\label{fig:t2d-cd-uc-conditionals}
\end{figure}

Lastly, we pursue a number or regionally restricted analyses. For each trait and for each of the seven geographic/genetic clusters described by Rosenberg et al (2002) \cite{Rosenberg:2002ga}, we compute a region specific $Q_X$ statistic to get a sense for the extent to which global signals we detect can be explained by variation among populations with these regions, and to highlight particular populations and traits which may merit further examination as more association data becomes available. The results are reported in Table \ref{regional-table}. We also apply our conditional Z-score approach at two levels of population structure: first at the level of Rosenberg's geographic/genetic clusters, testing each cluster in turn for how differentiated it is from the rest of the world, and second at the level of individual populations. The regional level Z-scores are useful for identifying signals of selection acting over broad regional scale or on deeper evolutionary timescales, while the population specific Z-scores are useful for identifying very recent selection that has only impacted a single population. We generally employ these regional statistics as a heuristic tool to localize signatures of selection uncovered in global analyses, or in cases where there is no globally interesting signal, to highlight populations or regions which may merit further examination as more association data becomes available. The result of these analyses are depicted in Figures \ref{fig:height-skin-bmi-conditionals} and \ref{fig:t2d-cd-uc-conditionals}, as well as Tables \ref{cond-height-region}-\ref{cond-uc-ind}.

\paragraph{Height}
We first analyzed the set of 180 height associated loci identified by Lango Allen and colleagues \cite{LangoAllen2010}, which explain about 10.5\% of the total variance for height in the mapping population, or about 15\% of heritability \cite{Zaitlen:2013hj}. This dataset is an ideal first test for our methods because it contains the largest number of associations identified for a single phenotype to date, maximizing our power gain over single locus methods (Figure \ref{power-plot}). In addition, Turchin and colleagues \cite{Turchin:2012dya} have already identified a signal of pervasive weak selection at these same loci in European populations, and thus we should expect our methods to replicate this observation.

Of the 180 loci identified by Lango Allen and colleagues, 
161 were present in our HGDP dataset. We used
these 161 loci in conjunction with the allele frequency data from the
HGDP dataset to estimate genetic values for height in the 52 HGDP
populations. Although the genetic values are correlated with the
observed heights in these populations, they are unsurprisingly 
imperfect predictions (see Figure \ref{Supp_fig_height_Gustafsson} and Table \ref{Gustafsson-height-table}, 
which compares our estimated genetic values to observed sex-average heights 
for the subset of HGDP populations with a close proxy in the 
dataset of Gustafsson and Lindenfors (2009)\cite{Gustafsson2009}). 

\begin{figure}[!ht]
	\includegraphics[width = \textwidth]{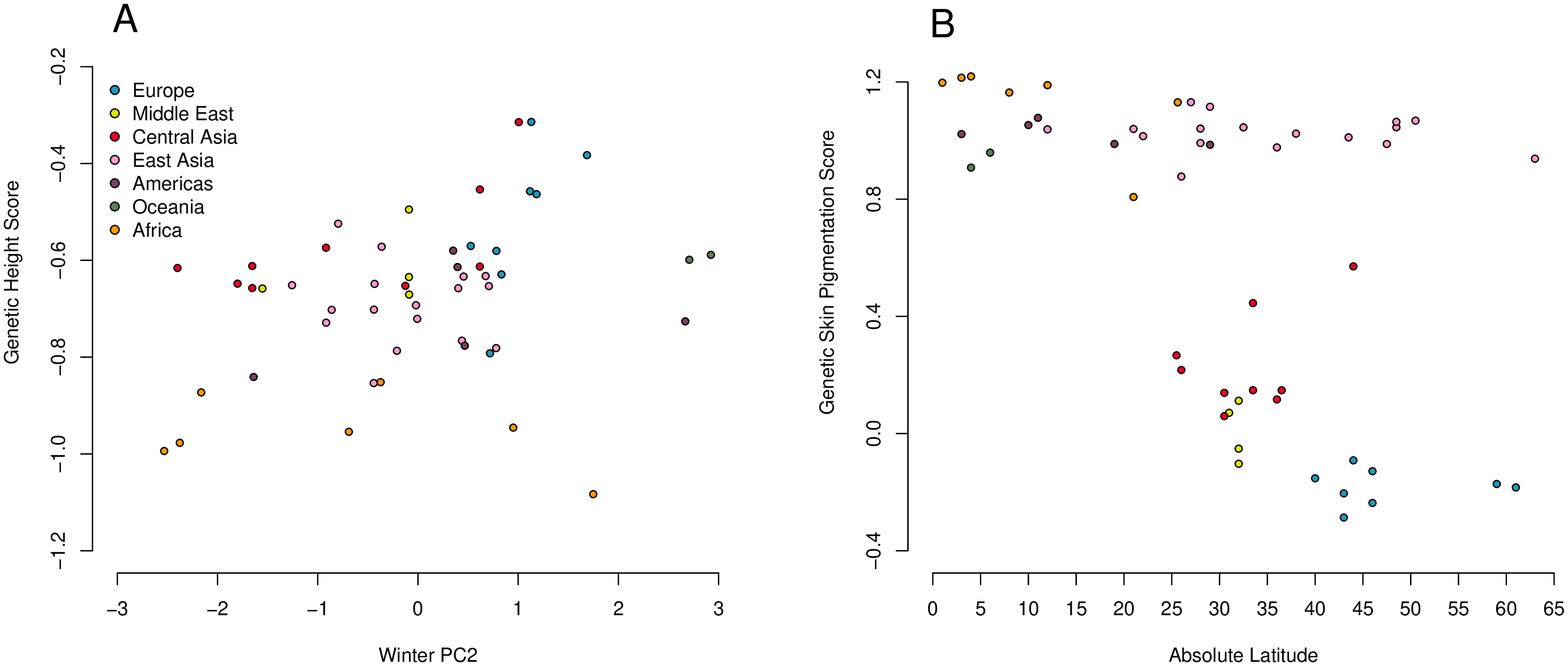}
	\caption{{\bf Estimated genetic height (A) and skin pigmentation score (B) plotted against winter PC2 and absolute latitude respectively.} Both correlations are significant against the genome wide background after controlling for population structure (Table \ref{results-table}). }
	\label{height-plus-skin}
\end{figure}

We find a signal of excessive correlation with winter PC2 (Figure \ref{height-plus-skin} and Table \ref{results-table}), 
but find no strong correlations with any other climatic variables.
Our $Q_X$ test also strongly rejects the neutral hypothesis, suggesting that our estimated genetic values 
are overly dispersed compared to the null model of neutral genetic
drift and shared population history (Figure \ref{QxHistograms} and
Table \ref{results-table}). These results are consistent with 
with directional selection acting in concert on alleles influencing
height to drive differentiation among populations at the level of the phenotype.




We followed up on these results by conducting regional level analyses, 
which indicate that our signal of excess variance arises primarily from extreme differentiation among populations within Europe (Table \ref{regional-table}).
Analyses using the conditional multivariate normal model indicate that this signal is driven largely by divergence between the French and Sardinian populations,
in line with Turchin et al's (2012) previous observation of a North-South gradient of height associated loci in Europe. 
We also find weaker signals of over-dispersion in other regions, but the globally significant $Q_X$ statistic can be 
erased by removing either the French or the Sardinian population from the analysis,
suggesting that the signal is primarily driven by differentiation among those two populations.



\paragraph{Skin Pigmentation}
We next analyzed data from a recent GWAS for skin pigmentation in an African-European admixed population of Cape Verdeans \cite{Beleza:2013cfa}, which identified four loci of major effect that explain approximately 35\% of the variance in skin pigmentation in that population after controlling for admixture proportion. 
Beleza et al (2013) report effect sizes in units of modified melanin (MM) index, 
which is calculated as $100\times \text{log} (1/\% \text{melanin reflectance at 650 nM})$, 
i.e. a higher MM index corresponds to darker skin, and a lower value to lighter skin.

We used these four loci to calculate a genetic skin pigmentation score in each of the HGDP populations. 
As expected, we identified a strong signal of excess variance among
populations, as well as a strong correlation with latitude (Figure
\ref{height-plus-skin} and Table \ref{results-table}), again consistent with
directional selection having acted on the phenotype of skin pigmentation to drive divergence among populations.
Note, however, that this signal was driven entirely by the fact that populations of western Eurasian descent have a lower genetic skin pigmentation score than populations of African descent.
Using only the markers from \cite{Beleza:2013cfa}, light skinned populations in East Asian and the Americas have a genetic skin pigmentation score
that is almost as high (dark) as that of most African populations, an effect that is clearly visible when we plot the measured skin pigmentation and skin reflectance of HGDP populations \cite{Jablonski:2000ds,Lao:2007jf} against their genetic values (see Figures \ref{Supp_fig_skin_pigmentation_Biasutti} and \ref{Supp_fig_skin_pigmentation_JablonskiChaplin}). The correlation with latitude is thus weaker than one might expect, given the known phenotypic distribution of skin pigmentation in human populations \cite{Jablonski:2000ds,Jablonski:2010hl}. To illustrate this point further, we re-ran the analysis on a subsample of the HGDP consisting of populations from Europe, the Middle East, Central Asia, and Africa. In this subsample, the correlation with latitude, and signal of excess variance, was notably stronger ($r^2=0.2$, $p=0.019$; $Q_X = 60.1$, $p=8\times 10^{-4}$).

This poor fit to observed skin pigmentation is due to the fact that we have failed to capture all of the loci that contribute to variation in skin pigmentation across the range of populations sampled, likely due to the partial convergent evolution of light skin pigmentation in Western and Eastern Eurasian populations \cite{Norton:2006kp}. 
Including other loci putatively involved in skin pigmentation \cite{Miller:2007jk,Edwards:2010kd} 
decreases the estimated genetic pigmentation score of the other Eurasian populations (Figures \ref{Supp_fig_skin_pigmentation_Biasutti} and \ref{Supp_fig_skin_pigmentation_JablonskiChaplin} and Table \ref{Supp_table_skin_pigmentation}),
but we do not include these in our main analyses as they differ in ascertainment (and the role of KITLG in pigmentation variation has been contested by \cite{Beleza:2013cfa}).

Within Africa, the San population has a decidedly lower genetic skin pigmentation score than any other HGDP African population.
This is potentially in accordance with the observation that the San
are more lightly pigmented than other African populations represented
by the HGDP \cite{Jablonski:2000ds} and the observation that other putative light skin pigmentation alleles have higher frequency in the San 
than other African populations  \cite{Norton:2006kp}. Although
there is still much work to be done on the genetic basis of skin
pigment variation within Africa, in this dataset a regional analysis of the six African populations alone identifies a marginally significant correlation with latitude ($r^2 = 0.62$, $p = 0.0612$), 
and a signal of excess variance among populations ($Q_X = 16.19$, $p =
0.01$), suggesting a possible role for selection in the shaping of
modern pigmentary variation within the continent of Africa.

\paragraph{Body Mass Index}
We next investigate two traits related to metabolic phenotypes (BMI and Type 2 diabetes), as there is a long history of adaptive hypotheses put forward to explain phenotypic variation among populations, with little conclusive evidence emerging thus far. We first focus on the set of 32 BMI associated SNPs identified by Speliotes and colleagues \cite{Speliotes2010} in their Table 1, which explain approximately 1.45\% of the total variance for BMI, or about 2-4\% of the additive genetic variance. Of these 32 associated SNPs, 28 were present in the HGDP dataset, which we used to calculate a genetic BMI score for each HGDP population.
We identified no significant signal of selection at the global level (Table \ref{results-table}).


Our regional level analysis indicated that the mean genetic BMI score is significantly lower that expected in East Asia ($Z = -2.48,\ p = 0.01$; see also Figure \ref{fig:height-skin-bmi-conditionals} and Table \ref{cond-bmi-region}), while marginal $Q_X$ statistics identify excess intraregional variation within East Asia and the Americas (Table \ref{regional-table}). While these results are intriguing, given the small fraction of the additive genetic variance explained by the ascertained SNPs and the lack of a globally significant signal or a clear ecological pattern or explanation, it is difficult to draw strong conclusions from them. For this reason BMI and other related traits will warrant reexamination as more association results arise and methods for analyzing association results from multiple correlated traits are developed.

\paragraph{Type 2 Diabetes}

We next investigated the 65 loci reported by Morris and colleagues \cite{Morris:2012iu} as associated with T2D, which explain $5.7\%$ of the total variance for T2D susceptibility, or about 8-9\% of the additive genetic variance. Of these 65 SNPs, 61 were present in the HGDP dataset. We used effect sizes from the stage 1 meta-analysis, and where a range of allele frequencies are reported (due to differing sample frequencies among cohorts), we simply used the average. Where multiple SNPs were reported per locus we used the lead SNP from the combined meta-analysis. Also note that Morris and colleagues report effects in terms odds ratios (OR), which can be converted into additive effects by taking the logarithm (the same is true of the IBD data from \cite{paper:2013iua}, analyzed below).

The distribution of genetic T2D risk scores showed no significant correlations with any of the five eco-geographic axes we tested, and was in fact fairly underdispersed worldwide relative to the null expectation due to population structure (Table \ref{results-table}).

Our regional level analysis revealed that while T2D genetic risk is well explained by drift in Africa, Central and Eastern Asia, Oceania, and the Americas, European populations have far lower T2D genetic risk than expected ($Z = -2.79$, $p = 0.005$) and Middle Eastern populations a higher genetic risk than expected ($Z = 2.37$, $p = 0.018$). It's not clear, however, that these observations should be interpreted as evidence for selection either in Europe or the Middle East. While the dichotomous regional labels ``Europe'' vs.\ ``Middle East'' explains the majority of the variance not accounted for by population structure ($r^2 = 0.77, p =5\times 10^{-4}$), this is essentially the same signal detected by the regional Z scores, and our $Q_X$ statistic finds no signal of excess variance ($Q_X = 10.9$, $p = 0.48$) among the twelve HGDP populations in these two regions. Expanding to the next most closely related region, we tested for a signal of excess differentiation between Central Asia and either Europe or Middle East, but find no convincing signal in either case ($r^2 = 0.13,\ p = 0.21;\ Q_X = 12.0,\ p = 0.75$ and $r^2 = 0.15,\ p = 0.19;\ Q_X = 9.8,\ p = 0.63$ respectively).
To the extent that our results are consistent with an impact of selection on the genetic basis of T2D risk, they appear to be consistent primarily with a scenario in which selection has pushed the frequency of alleles that increase T2D risk up in Middle Eastern populations and down in European populations. This is an extremely subtle signal, which arises only after deep probing of the data, and as such we are skeptical as to whether our results represent a meaningful signal of selection.

A number of investigators have claimed that individual European GWAS loci for Type 2 Diabetes
show signals of selection \cite{Helgason:2007gz,Hancock:2008cn,Hancock:2010gi,Klimentidis:2010bu}, 
a fact that is seen as support for the idea 
that genetic variation for T2D risk has been shaped by local adaptation,
potentially consistent with a variation on the thrifty genotype hypothesis \cite{Neel:1962tj}.
However, our result suggest that local adaptation has not had a large role in shaping
the present day world-wide distribution of T2D susceptibility alleles (as mapped to date in Europe).
One explanation of this discrepancy is that it is biologically unrealistic that the 
phenotype of T2D susceptibility would exhibit strong adaptive differentiation. 
Rather, local adaptation may have shaped some pleiotropically related phenotype (which shares only some of the loci involved).
However, as seen in Figure \ref{power-plot}, our methods have better power than single locus statistics
so long as there is a reasonable correlation ($\phi > 0.3$) between the focal phenotype and the one under selection.
As such, the intersection of our results with previous studies support the idea that local adaptation has had little direct influence on 
the genetic basis of T2D or closely correlated phenotypes, but that a handful of individual SNPs associated with 
T2D may have experienced adaptive differentiation as a result of their
function in some other phenotype.


\

\paragraph{IBD} Finally, we analyzed the set of associations reported for Crohn's Disease (CD) and Ulcerative Colitis (UC) \cite{paper:2013iua}. Because CD and UC are closely connected phenotypes that share much of their genetic etiology, Jostins and colleagues used a likelihood ratio test of four different models (CD only, UC only, both CD and UC with equal effects on each, both CD and UC with independent effects) to distinguish which SNPs where associated with either or both phenotypes, and to assign effect sizes to SNPs (see their supplementary methods section 1d). We take these classifications at face value, resulting in two partially overlapping lists of 140 and 135 SNPs associated with CD and UC, which explain $13.6\%$ and $7.5\%$ of disease susceptibility variance respectively. Of these, there are 95 SNPs for CD and 89 SNPs for UC were present in our HGDP dataset, and these remaining SNPs on which our analyses are based explain 9\% and 5.1\% of the total variance. For now, we treat these sets of loci independently, and leave the development of methods that appropriately deal with correlated traits for future work.

\begin{figure}[!ht]
	\includegraphics[width = \textwidth]{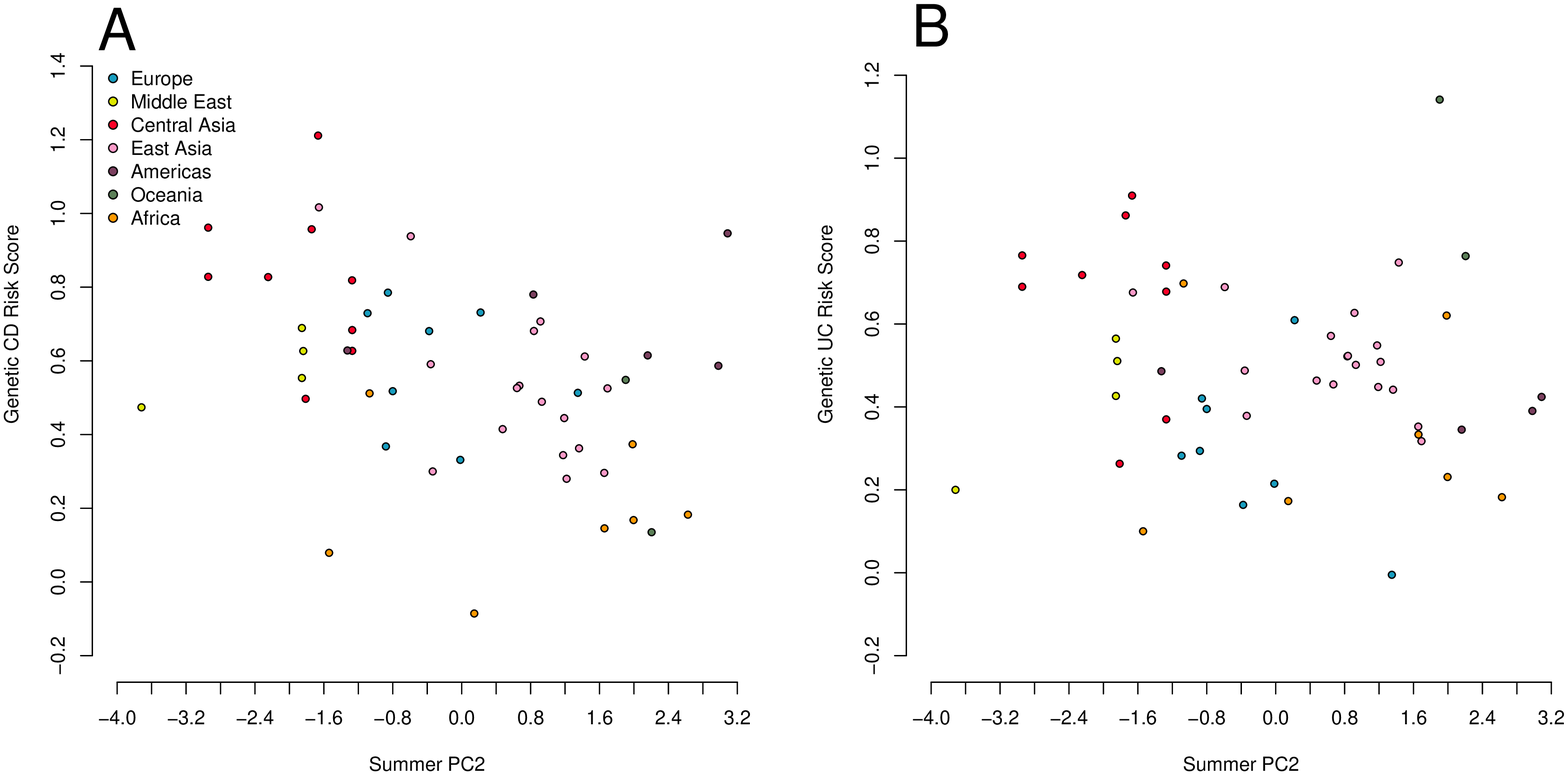}
	\caption{{\bf Estimated genetic risk score for Crohn's Disease (A) and Ulcerative Colitis (B) risk plotted against summer PC2.} Both correlations are significant against the genome wide background after controlling for population structure (Table \ref{results-table}). Since a large proportion of SNPs underlying these traits are shared, we note that these results are not independent.}\label{fig:uc-plus-cd}
\end{figure}

We used these sets of SNPs to calculate genetic risk scores for CD and UC across the 52 HGDP populations. Both CD and UC showed strong negative correlations with summer PC2 (Figure \ref{fig:uc-plus-cd}), while CD also showed a significant correlation with winter PC1, and a marginally significant correlation with summer PC1 (Table \ref{results-table}). 

We did not observe any significant $Q_X$ statistics for either trait, either at the global or the regional level, suggesting that our environmental correlation signals most likely arise from subtle differences between regions, as opposed to divergence among closely related populations. Indeed, we find moderate signals of regional level divergence in Europe (UC: $Z = -2.08, p = 0.04$), Central Asia (CD: $Z = 2.21, p = 0.03$), and East Asia (CD: $Z = -1.90, p = 0.06$ and UC: $Z = -2.12, p = 0.03$; see also Figure \ref{fig:t2d-cd-uc-conditionals} and Tables \ref{cond-uc-region} and \ref{cond-uc-ind}).

%

\section*{Discussion}

In this paper we have developed a powerful framework for identifying the influence of local adaptation on the genetic loci underlying variation in polygenic phenotypes. Below we discuss two major issues related to the application of such methods, namely the effect of the GWAS ascertainment scheme on our inference, and the interpretation of positive results. 

\subsection*{Ascertainment and Population Structure}
Among the most significant potential pitfalls of our analysis (and the most likely cause of a false positive) is the fact that 
the loci used to test for the effect of selection on a given phenotype have 
been obtained through a GWAS ascertainment procedure, 
which can introduce false signals of selection if potential confounds are not properly controlled. 
We condition on simple features of the ascertainment process via our allele matching procedure, 
but deeper issues may arise from artifactual associations that result from the effects of population structure in the GWAS ascertainment panel.
Given the importance of addressing this issue to the broader GWAS community,
a range of well developed methods exist for doing GWAS in structured populations,  
and we refer the reader to the existing literature for a full discussion \cite{Freedman:2004dk,Campbell:2005bt,Price:2006cd,Kang:2008bx,Price2010,Diao:2012bo,Liu:2013ee}. Here, we focus on two related issues. First, the propensity of population structure in the GWAS ascertainment panel to generate false positives in our selection analysis, and second, the difficulties introduced by the sophisticated statistical approaches employed to deal with this issue when GWAS are done in strongly structured populations.

The problem of population structure arises generally when there is a correlation in the ascertainment panel between phenotype and ancestry such that SNPs that are ancestry informative will appear to be associated with the trait, even when no causal relationship exists \cite{Campbell:2005bt}. This phenomenon can occur regardless of whether the correlation between ancestry and phenotype is caused by genetic or environmental effects. To make matters worse, multiple false positive associations will tend to line up with same axis of population structure. If the populations being tested with our methods lie at least partially along the same axis of structure present in the GWAS ascertainment panel, then the ascertainment process will serve to generate the very signal of positive covariance among like effect alleles that our methods rely on to detect the signal of selection.

The primary takeaway from this observation is that the more diverse the array of individuals sampled for a given GWAS are with respect to ancestry, 
the greater the possibility that failing to control for population
structure will generate false associations (or bias effect sizes) and hence false positives for our method.


What bearing do these complications have on our empirical results? 
The GWAS datasets we used can be divided into those conducted within populations of European descent 
and the skin pigmentation dataset (which used an admixed population). 
We will first discuss our analysis of the former. 

The European GWAS loci we used were found in relatively homogeneous populations, 
in studies with rigorous standards for replication and control for population structure. 
Therefore, we are reasonably confident that these loci are true positives.
Couple this with the fact that they were ascertained in populations that are fairly homogenous relative to the global scale of our analyses,
and it is unlikely that population structure in the ascertainment panels is driving our positive signals.
One might worry that we could still generate false signals by including European populations in our analysis, 
however many of the signals we see are driven by patterns outside of Europe (where the influence of structure within Europe should be much lessened).
For height, where we do see a strong signal from within Europe, 
we use a set of loci that
have been independently verified using a family based design that is immune to the effects of population structure \cite{Turchin:2012dya} . 


We further note that for a number of GWAS datasets, including some of those analyzed here, studies of non-European populations have replicated many of the loci identified in European populations \cite{Cho:2009hk,Cho:2011jk,Voight:2010il,Kooner:2011hi,NDiaye:2011ur,Carty2012,Monda:2013ipa}, and for many diseases, the failure of some SNPs to replicate, as well as discrepancies in effect size estimate, are likely due to simple considerations of statistical power and differences in patterns of LD across populations \cite{Carlson:2013gn,Marigorta:2013cr}. This suggests that, at least for GWAS done in relatively homogenous human populations, structure is unlikely to be a major confounding factor.


The issue of population structure may be more profound for our style of approach
when GWAS are conducted using individuals from more strongly structured populations. 
In some cases it is desirable to conduct GWAS in such populations as locally adaptive alleles will be present at intermediate frequencies in these broader samples, whereas they may be nearly fixed in more homogeneous samples. 
A range of methods have been developed to adjust for population structure in these setting \cite{Kang:2010fg,Zhou:2012gh,Liu:2013hg}. While generally effective in their goal, these methods present their own issues for our selection analysis. Consider the extreme case, such as that of Atwell et al (2010) \cite{Atwell2010}, who carried out a GWAS in \textit{Arabidopsis thaliana} for 107 phenotypes across an array of 183 inbred lines of diverse geographical and ecological origin. Atwell and colleagues used the genome-wide mixed model program EMMA \cite{Kang:2008bx,Kang:2010fg,Zhou:2012gh} to control for the complex structure present in their ascertainment panel. This practice helps ensure that many of the identified associations are likely to be real, 
but also means that the loci found are likely to have unusual frequencies patterns across the species range. 
This follows from the fact that the loci identified as associated with the trait must stand out as being correlated with the trait in a way not predicted by the individual kinship matrix (as used by EMMA and other mixed model approaches). Our approach is predicated on the fact that we can use genome-wide patterns of kinship to adjust for population structure, but this correction is exactly the null model that loci significantly associated with phenotypes by mixed models have overcome. 
For this reason, both the theoretical $\chi^2$ distribution of the $Q_X$ statistic, as well as the empirical null distributions we construct from resampling, may be inappropriate. 

The Cape Verde skin pigmentation data we used may qualify as this second type of study. 
The Cape Verde population is an admixed population of African/European descent, 
and has substantial inter-individual variation in admixture proportion.  
Due to its admixed nature, the population segregates alleles which would not 
be at intermediate frequency in either parental population, 
making it an ideal mapping population.

Despite the considerable population structure, the fact that intermarriage 
continues to mix genotypes in this population means that much of the LD due to the African/European population structure has been broken up (and the remaining LD is well predicted 
by an individual's genome-wide admixture coefficient). 
Population structure seems to have been well controlled for in this study, 
and a number of the loci have been replicated in independent admixed populations. 
While we think it unlikely that the four loci we use are false associations, 
they could in principle suffer from the structured ascertainment issues described above,
so it is unclear that the null distributions we use are strictly appropriate.
That said, provided that Beleza and colleagues have  appropriately controlled for population structure, 
under neutrality there would be no reason to expect that the correlation among the loci should be strongly positive with respect to the sign of their effect on the phenotype,
and thus the pattern observed is at least consistent with a history of selection,
especially in light of the multiple alternative lines of evidence for adaptation on the basis of skin pigmentation \cite{Jablonski:2000ds,Sabeti:2007uv,Lao:2007jf,Williamson:2007db,Sturm2009,Jablonski:2010hl}.

Further work is needed to determine how best to modify the tests proposed herein to deal 
with GWAS performed in structured populations.

\subsection*{Complications of Intepretation}
Our understanding of the genetic basis of variation in complex traits
remains very incomplete, and as such the results of these analyses
must be interpreted with cautiously. That said, because our methods are
based simply on the rejection of a robust, neutral null model, an incomplete knowledge of the genetic basis of a given trait should only lead to a loss of statistical power, and not to a high false positive rate.
	
For all traits analyzed here except for skin pigmentation, the within population variance for genetic value is considerably larger than the variance between populations. 
This suggests that much of what we find is relatively subtle adaptation even on the level of the phenotype,
and emphasizes the fact that for most genetic and phenotypic variation in humans, the majority of the variance 
is within populations rather than between populations (see Figures \ref{ind-var-height}--\ref{ind-var-UC}).
In many cases, the influence of the environment likely plays a stronger role in the differences between populations for true phenotypes
than the subtle differences we find here (as demonstrated by the rapid change in T2D incidence with changing diet, e.g. \cite{Franco:2013hb}). 
That said, an understanding of how adaptation has shaped the genetic basis 
of a wide variety of phenotypes is clearly of interest, 
even if environmental differences dominate as the cause of present day population differences,
as it informs our understanding of the biology and evolutionary history of these traits.

The larger conceptual issues relate to the interpretation of our positive findings, which we detail below.
A number of these issues are inherent to the conceptual interpretation of evidence for local adaptation \cite{Kawecki:2004hx}.

\paragraph{Effect Size Heterogeneity and Misestimation} 
In all of our analyses, we have simply extrapolated GWAS effect sizes measured in one population and one environment to the entire panel of HGDP populations. It is therefore prudent to consider the validity of this assumption, as well as the implications for our analyses when it is violated. Aside from simple measurement error, there are two possible reasons that estimated effect sizes from GWAS may not reflect the true effect sizes. 


The first is that most GWAS hits likely identify tag SNPs that are in
strong LD with causal sites that are physically nearby on the
chromosome, rather than actual causal sites themselves
\cite{Marigorta:2013cr,Carlson:2013gn}. 
This acts to reduce the estimated effect size in the GWAS sample. More
importantly for the interpretation of our signals, patterns of LD between tag
SNPs and causal sites will change over evolutionary time, 
and so a tag SNP's allele frequencies will be an imperfect measure of
the differentiation of the causal SNP over the sampled populations. This should lead to a reduction in our power to detect the effect of selection in much the same way that power is reduced when selection acts on a trait that is genetically correlated with the trait of interest (Figure 1B). This effect will be especially pronounced when the populations under study have a shorter scale of LD than the populations in which the effect have been mapped (e.g. when applying effect sizes estimated in Europe to population of African descent). In the case that selection has not affected the trait of interest, the effect sizes have no association whatsoever with the distribution of allele frequencies across populations unless such an association is induced by the ascertainment process, as described above. Therefore, changes in the patterns of LD between identified tag SNPs and causal sites will not lead to an excess of false positives if the loci under study have not been subject to spatially varying selection pressures.

The second is that the actual value of the additive effect at a causal
site may change across environments and genetic backgrounds due to
genotype-by-genotype (i.e. functional epistasis) and
genotype-by-environment interactions. Although the response at a given
locus due to selection depends only the additive effect of the allele
in that generation, the additive effect itself is a function of the
environment and the frequencies of all interacting loci. As all of
these can change considerably during the course of evolution, the
effects estimated in one population may not apply in other
populations, either in the present day, or over history of the populations \cite{Fisher:1918wp,Wade:2002ul}. We first wish to stress that, as above, because our tests rely on rejection of a null model of drift, differences in additive effects among populations or over time will not lead to an excess of false positives, provided that the trait is truly neutral. Such interactions can, however, considerably complicate the interpretation of positive results. For example, different sets of alleles could be locally selected to maintain a constant phenotype across populations due to gene-by-environment interactions. Such a scenario could lead to a signal of local adaptation on a genetic level but no change in the phenotype across populations, a phenomenon known as countergradient variation \cite{Conover:1995dt}. 

It will be very difficult to know how reasonable it is to extrapolate
effect sizes among populations without repeating measurements in
different populations and different environments. 
Perhaps surprisingly, the existing evidence suggests that for a variety of highly polygenic
traits, effects sizes and directions may be surprisingly consistent
across human populations \cite{Cho:2009hk,Cho:2011jk,Voight:2010il,Kooner:2011hi,NDiaye:2011ur,Carty2012,Monda:2013ipa,Carlson:2013gn,Marigorta:2013cr}. There is no particular reason to believe that this will hold as a general rule across traits or across species, and thus addressing this issue will require a great deal more functional genetic work and population genetic method development, a topic which we discuss briefly below in Future Directions.

%
%
%
%
%

\paragraph{Missing variants}
As the majority of GWAS studies are performed in a single population
they will often miss variants contributing to phenotypic
variation. This can occur due to GxG or GxE interactions as outlined above, but
also simply because those variants are absent (or at low frequency)
due to drift or selection among the populations.
Such cases will not create a false signal of selection if only drift is involved,
however, they do complicate the interpretation of positive signals. 
A particularly dramatic example of this is offered by our analysis of skin pigmentation associated loci,
whose frequencies are clearly shaped by adaptation. 
The alleles found by a GWAS in the Cape Verde population completely fail to predict
the skin pigmentation of East Asians and Native Americans. 
This reflects the fact that a number of the alleles responsible for light skin pigmentation in those populations
are not variable in Cape Verde due to the partially convergent
adaptive evolution of light skin pigmentation \cite{Norton:2006kp}.
As a result, when we take the Eurasian HGDP populations 
we see a significant correlation between genetic skin pigmentation score and longitude ($r^2 = 0.15, p = 0.015$), 
despite the fact that no such phenotypic correlation exists. 
While the wrong interpretation is easy to avoid here because we have a good
understanding of the true phenotypic distribution, 
for the majority of GWAS studies such complications will be subtler 
and so care will have to be taken in the interpretation of positive results.



\paragraph{Loss of constraint and mutational pressure}
One further complication in the interpretation of our results is 
in how loss of constraint may play a role in driving apparent signals of 
local adaptation. Traits evolving under uniform stabilizing selection across all populations 
should be less variable than predicted by our covariance model of drift, due to negative covariances among loci, 
and so should be underrepresented in the extreme tails of our environmental correlation statistics
and the upper tail of $Q_X$. As such, 
loss of constraint (i.e.\ weaker stabilizing selection in some populations than others), should not on its own 
create a signal of local adaptation. While the loci underpinning the 
phenotype can be subject to more drift in those populations, 
there is no systemic bias in the direction of this drift. 
Loss of constraint, therefore, will not tend to create significant
environmental correlations or systematic covariance between alleles of
like effect.

An issue may arise, however, when loss of constraint is paired with biased mutational input 
(i.e.\ new mutations are more likely to push the phenotype in one direction than another \cite{Zhang:2008fi})
or asymmetric loss of constraint (selection is relaxed on one tail of the phenotypic
distribution). 
Under these two scenarios, alleles that (say) increase the phenotype
would tend to drift up in frequency in the populations with loss of constraint, 
creating systematic trends and positive covariance among like effect alleles at different loci, 
and resulting in a positive signal under our framework.
While one would be mistaken to assume that the signal was necessarily that of recent positive directional selection, these scenarios do still imply that selection pressures on the genetic basis of the phenotype vary across space. 
Positive tests under our methods are thus fairly robust in being signals of differential selection among populations,
but are themselves agnostic about the specific processes involved.
Further work is needed to establish whether these scenarios
can be distinguished from recent directional selection based on only allele frequencies and effect sizes, 
and as always, claims of recent adaptation should be supported by multiple lines of evidence 
beyond those provided by population genomics alone.

\paragraph{Future directions}
 In this article we have focused on methods development and so have not fully explored the range of populations and phenotypes to which our methods could be applied.
Of particular interest is the possibility of applying these methods to GWAS performed in other species 
where the ecological determinants of local adaptation are better understood \cite{Atwell2010,Fournier-Level2011}.

One substantial difficulty with our approach, 
particularly in its application to other organisms, 
is that genome-wide association studies 
of highly polygenic phenotypes require very large sample sizes to map even a fraction of the total genetic variance. 
One promising way to partially sidestep this issue 
is by applying methods recently developed in animal and plant breeding.
In these genomic prediction/selection approaches, one does not attempt to map individual markers,
but instead concentrates on predicting an individual's genetic value for a given phenotype using all markers simultaneously \cite{Meuwissen2001,Hayes:2009ec,Meuwissen:2013bb}. 
This is accomplished by fitting simple linear models to genome-wide genotyping data, in principle allowing common SNPs to tag the majority of causal sites throughout the genome without attempting to explicitly identify them \cite{Zhou:2013fh}.
These methods have been applied to a range of species, including humans \cite{Yang:2010kn,Davies:2011dn,Yang:2011cb,Lee:2012iu,delosCampos:2010ho,delosCampos:2012ku,delosCampos:2013fd}, demonstrating that these predictions can potentially explain a relatively high fraction of the additive genetic variance within a population
(and hence much of the total genetic variance).
As these predictions are linear functions of genotypes, and hence allele frequencies, 
we might be able to predict the genetic values of sets of closely related populations 
for phenotypes of interest and apply very similar methods to those developed here.
Such an approach may allow for substantial gains in power,
as it would greatly increase the fraction of the genetic variance used in the analyses.
However, if the only goal is to establish evidence for 
local adaptation in a given phenotype, then because measurements of true phenotypes inherently include all of the underlying loci, the optimal approach is to perform a common garden experiment
and employ statistical methods such as those developed by Ovaskainen and colleagues \cite{Ovaskainen:2011ku,Karhunen:2012dy,Karhunen:2013eg}, 
assuming such experiments can be done.

As discussed in various places above, it is unlikely that all of the loci underpinning the 
genetic basis of a trait will have been subject to the same selection pressures, 
due to their differing roles in the trait and their pleiotropic effects.
One potential avenue of future investigation is whether, given a large set of loci involved 
in a trait, we can identify sets of loci in particular pathways or with a particular set of functional attributes 
that drive the signal of selection on the additive genetic basis of a trait.

Another promising extension of our approach is to deal explicitly with multiple correlated phenotypes. 
With the increasing number of GWAS efforts both empirical and methodological work 
are beginning to focus on understanding the shared genetic basis of various phenotypes \cite{paper:2013iua,GlobalLipidsGeneticsConsortium:2013hs}.
This raises the possibility that we may be able to disentangle the genetic
basis of which phenotypes are more direct targets of selection, and 
which are responding to correlated selection on these direct targets 
(for progress along these lines using $Q_{ST}$, see \cite{Kremer:1997vf,BLOWS:2007jo,Chenoweth:2008gv,Ovaskainen:2011ku}). 
Such tools may also offer a way of incorporating GxE interactions, 
as multiple GWAS for the same trait in different environments can be
treated as correlated traits \cite{Falconer:1952uz}.


As association data for a greater variety of populations, species, and traits becomes available, we view the methods described out here as a productive way forward in developing a quantitative framework to explore the genetic and phenotypic basis of local adaptation.

\section*{Materials and Methods}

\subsection*{Mean Centering and Covariance Matrix Estimation}
Written in matrix notation, the procedure of mean centering the estimated genetic values and dropping one population from the analysis can be expressed as
\begin{equation}
	\vec{Z^\prime} = \mathbf T \vec{Z}
\end{equation}
where $\mathbf T$ is an $M -1$ by $M$ matrix with $\frac{M-1}{M}$ on the main diagonal, and $-\frac{1}{M}$ elsewhere.

\label{mean-centering}
In order to calculate the corresponding expected neutral covariance structure about this mean, we use the following procedure. Let $\mathbf G$ be an $M$ by $K$ matrix, where each column is a vector of allele frequencies across the $M$ populations at a particular SNP, randomly sampled from the genome according to the matching procedure described below. 
Let $\epsilon_k$ and $\epsilon_i$ be the mean allele frequency in columns $k$ and $i$ of $\mathbf G$ respectively, and let $\mathbf S$ be a matrix such that $s_{ki} = \frac{1}{\sqrt{\epsilon_k(1-\epsilon_k)\epsilon_i(1-\epsilon_i)}}$.
With these data, we can estimate $\mathbf{F}$ as
\begin{equation}
	\mathbf{F} =\mathbf{TG} \mathbf  S \mathbf{G^TT^T}.
	\label{cov-mat-estimation}
\end{equation}
This transformation performs the operation of centering the matrix at the mean value, and rooting the analysis with one population by dropping it from the covariance matrix (the same one we dropped from the vector of estimated genetic values), resulting in a covariance matrix describing the relationship of the remaining $M-1$ populations. This procedure thus escapes the singularity introduced by centering the matrix at the observed mean of the sample.

As we do not get to observe the population allele frequencies, the entries of $\mathbf{G}$ are the sample frequencies at the randomly chosen loci, and thus the covariance matrix $\mathbf{F}$ also includes the effect of finite sample size. Because the noise introduced by the sampling of individuals is uncorrelated across populations (in contrast to that introduced by drift and shared history), the primary effect is to inflate the diagonal entries of the matrix by a factor of $\frac{1}{n_m}$, where $n_m$ is the number of individuals sampled in population $m$ (see the supplementary material of \cite{pickrell:2012fu} for discussion). This means that our population structure adjusted statistics also approximately control for differences in sample size.

\paragraph{Standardized environmental variable}
Given a vector of environmental variable measurements for each population, we apply both the $\mathbf{T}$ and Cholesky tranformation as for the estimated genetic values
\begin{equation}
	\vec{Y^\prime} = \mathbf C^{-1} \mathbf T \vec{Y}. \label{stdize_environ}
\end{equation}
This provides us with a set of $M-1$ adjusted observations for the environmental variable which can be compared to the transformed genetic values for inference. This step is necessary as we have rotated the frame of reference of the estimated genetic values, and so we must do the same for the environmental variables to keep them both in a consistent reference frame.


\subsection*{Identifying Outliers with Conditional MVN Distributions}
As described in the Results, we can use our multivariate normal model of relatedness to obtain the expected distribution of genetic values for an arbitrary set of populations, conditional on the observed values in some other arbitrary set.

We first partition our populations into two groups, those for which we want to obtain the expected distribution of genetic values (group 1), and those on which we condition in order to obtain this distribution (group 2). We then re--estimate the covariance matrix such that it is centered on the mean of group 2. This step is necessary because the amount of divergence between the populations in group 1 and the mean of group 2 will always be greater than the amount of divergence from the global mean, even under the neutral model, and our covariance matrix needs to reflect this fact in order to make accurate predictions. We can obtain this re-parameterized $\mathbf{F}$ matrix as follows. If $M$ is the total number of populations in the sample, then let $q$ be the number of populations in group one, and let $M-q$ be the number of populations in group 2. We then define a new $\mathbf{T_R}$ matrix such that the $q$ columns corresponding the populations in group one have 1 on the diagonal, and 0 elsewhere, while the $M-q$ columns corresponding to group two have $\frac{M-q-1}{M-q}$ on the diagonal, and $-\frac{1}{M-q}$ elsewhere. We can then re--estimate a covariance matrix that is centered at the mean of the $M-q$ populations in group 2. Recalling our matrices $\mathbf{G}$ and $\mathbf{S}$ from \eqref{cov-mat-estimation}, this matrix is calculated as
\begin{align}
	\mathbf{F_R} = \mathbf{T_R}\mathbf{G}\mathbf{S}\mathbf{G^T}\mathbf{T_R^T}
\end{align}
where we write $\mathbf{F_R}$ to indicate that it is a covariance matrix that has been re-centered on the mean of group two.

Once we have calculated this re--centered covariance matrix, we can use well known results from multivariate normal theory to obtain the expected joint distribution of the genetic values for group one, conditional on the values observed in group two.

We partition our vector of genetic values and the re--centered covariance matrix such that
 \begin{align}
	\vec{X}	&=	\begin{bmatrix}
					\vec{X}_1 \\
					\vec{X}_2
				\end{bmatrix}\\
			&\text{and} \notag \\
			\mathbf{F}_{R} &= 	\begin{bmatrix}
							\mathbf{F}_{11}	&\mathbf{F}_{12} \\
							\mathbf{F}_{21}	&\mathbf{F}_{22}
						\end{bmatrix}
\end{align}
where $\vec{X}_1$ and $\vec{X}_2$ are vectors of genetic values in group 1 and 2 respectively, and $\mathbf{F}_{11}$, $\mathbf{F}_{22}$ and $\mathbf{F}_{12} = \mathbf{F}_{21}^T$ are the marginal covariance matrices of populations within group 1, within group 2, and across the two groups, respectively. Letting $\mu_1= \mu_2 = \frac{1}{M-q}\sum_{m = M-q}^M X_m$ (i.e.\ the sum of the elements of $\vec{X}_2$), we wish to obtain the distribution
\begin{align}
	\vec{X}_1 | \vec{X}_2, \mu_1, \mu_2  \sim MVN(\vec{\xi}, \mathbf{\Omega}),
\end{align}
where $\vec{\xi}$ and $\mathbf{\Omega}$ give the expected means and covariance structure of the populations in group 1, conditional on the values observed in group 2. These can be calculated as
\begin{align}
	\vec{\xi} = \mathbb{E}[\vec{X}_1 | \vec{X}_2, \mu_1, \mu_2] &= \mu_1\vec{1} + \mathbf{F}_{12}\mathbf{F}_{22}^{-1} \left(\vec{X}_2 - \mu_2\vec{1}\right) \label{eq:cond-mean} \\
	&\text{and} \notag\\
	\mathbf{\Omega} = \text{Cov}[X_1 | X_2, \mu_1, \mu_2]  &= \mathbf{F}_{11} -  \mathbf{F}_{12}\mathbf{F}_{22}^{-1}\mathbf{F}_{21}.
\end{align}
where the one vectors in line \eqref{eq:cond-mean} are of length $q$ and $M-q$ respectively. 

This distribution is itself multivariate normal, and as such this framework is extremely flexible, as it allows us to obtain the expected joint distribution for arbitrary sets of populations (e.g.\ geographic regions or continents), or for each individual population. Further,
\begin{align}
	\mathbb{E}\left[\frac{1}{q}\sum_{m=1}^q X_m \biggm | \vec{\xi},\mathbf{\Omega} \right] &= \frac{1}{q}\sum_{m=1}^q \xi_m \\
	&\text{and} \notag\\
	\text{Var}\left[\frac{1}{q}\sum_{m=1}^q X_m \biggm| \vec{\xi},\mathbf{\Omega}\right] &= \frac{V_A}{q^2}\sum_{m=1}^q\sum_{n=1}^q \mathbf{\omega}_{mn}.
\end{align}
where $\omega_{nm}$ denotes the elements of $\mathbf{\Omega}$. In words, the conditional expectation of the mean estimated genetic value across group 1 is equal to the mean of the conditional expectations, and its variance is equal to the mean value of the elements of the conditional covariance matrix. As such we can easily calculate a Z score (and corresponding p value) for group one as a whole as
\begin{align}
	Z = \frac{\frac{1}{q}\sum_{m=1}^q X_m- \frac{1}{q}\sum_{m=1}^q \xi_m}{\frac{1}{q}\sqrt{V_A\sum_{m=1}^q\sum_{n=1}^q \omega_{m,n}}}.
\end{align}
This Z score is a normal random variable with mean zero, variance one under the null hypothesis, and thus measures the divergence of the genetic values between the two populations relative to the null expectation under drift. Note that the observation of a significant Z score in a given population or region cannot necessarily be taken as evidence that selection has acted in that population or region, as selection in the some of the populations on which we condition (especially the closely related ones) could be responsible for such a signal. As such, caution is warranted when interpreting the output of these sort of analyses, and is best done in the context of more explicit information about the demographic history, geography, and ecology of the populations.

\subsection*{The Linear Model at the Individual Locus Level}

As with our excess variance test, explored in the main text, it is natural to ask how our environmental correlation tests can be written in terms of allele frequencies at individual loci.

As noted in \eqref{eq:transformed-freqs}, we can obtain for each underlying locus a set of transformed allele frequencies, which have passed through the same transformation as the estimated genetic values. We assume that each locus $\ell$ has a regression coefficient
\begin{equation}
	\beta_{\ell} = \gamma\alpha_{\ell}
\end{equation}
where $\gamma$ is shared across all loci so that
\begin{equation}
	p_{m \ell }' \sim \gamma\alpha_{\ell} Y_m' + e_{m\ell}
\end{equation}
where the $e_{m\ell}$ are independent and identically distributed residuals. 
We can find the maximum likelihood estimate $\hat{\gamma}$ by treating $\alpha_{\ell} Y_m^\prime$ as the linear predictor, 
and taking the regression of the combined vector $\vec{p^\prime}$, across all populations and loci, on the combined vector $\overrightarrow{\alpha Y^\prime}$. As such
\begin{equation}
	\hat{\gamma} = \frac{Cov(p^\prime,\alpha Y')}{Var(\alpha Y')}
\end{equation}
we can decompose this into a sum across loci such that
\begin{equation}
	\hat{\gamma} = \frac{\frac{1}{L} \sum_{\ell} Cov(p_{\ell}',\alpha_{\ell} Y')}{\frac{1}{L} \sum_{\ell} Var(\alpha_{\ell} Y)} = \frac{1}{\sum_{\ell} \alpha_{\ell}^2} \frac{\sum_{\ell} \alpha_{\ell} Cov(p_{\ell}',Y')}{Var(Y')}.  \label{slope_mean_locus}
\end{equation}
As noted in \eqref{eq:transformed-freqs}, our transformed genetic values can be written as
\begin{equation}
	X_m = 2 \sum_{\ell} \alpha_{\ell} p_{m\ell}'
\end{equation}
and so the estimated slope ($\hat{\beta}$) of our regression ($\vec{X} = \beta \vec{Y'} + \vec{e}$) is 
\begin{equation}
	\hat{\beta} = \frac{Cov(X,Y')}{Var(Y)} = \frac{2 \sum_{\ell} \alpha_{\ell} Cov(p_{\ell}',Y')}{Var(Y')} \label{slope_breeding_value}
\end{equation}
Comparing these equations, the mean regression coefficient at the individual loci \eqref{slope_mean_locus}
and the regression coefficient of the estimated genetic values \eqref{slope_breeding_value} 
are proportional to each other via a constant that is given by one over two times the sum of the effect sizes squared (i.e. $\gamma = \frac{1}{2\sum_{\ell} \alpha_{\ell}^2} \beta$). Our test based on estimating the regression of genetic values on the environmental variable is thus mathematically equivalent to an approach in which we assume that the regression coefficients of individual loci on the environmental variable are proportional to one another via a constant that is a function of the effect sizes. Such a relationship can also be demonstrated for the correlation coefficient ($r^2$) calculated at the genetic value level and at the individual locus level (this is not necessarily true for the rank correlation $\rho$), however the algebra is more complicated, and thus we do not show it here.

This is in contrast to the $r^2$ \textit{enrichment} statistic we compute for the power simulations, in which we assume that the correlations of individual loci with the environmental variable are independent of one another, and then perform a test for whether more loci individually show strong correlations with the environmental variable than we would expect by chance.

%

\subsection*{HGDP data and imputation}

We used imputed allele frequency data in the HGDP, where the imputation was performed as part of the phasing procedure of \cite{Pickrell:2009if}, as per the recommendations of \cite{Conrad2006}. We briefly recap their procedure here:

Phasing and imputation were done using fastPHASE \cite{Scheet:2006he}, 
with the settings that  allow variation in the switch rate between subpopulations. 
The populations were grouped into subpopulations corresponding to the clusters identified in \cite{Rosenberg:2002ga}. 
Haplotypes from the HapMap YRI and CEU populations were included as known, as they were phased in trios and are highly accurate. 
HapMap JPT and CHB genotypes were also included to help with the phasing. 

\subsection*{Choosing null SNPs}
\label{Methods:choosing-null-snps}
Various components of our procedure involve sampling random sets of SNPs from across the genome. While we control for biases in our test statistics introduced by population structure through our $\mathbf{F}$ matrix, we are also concerned that subtle ascertainment effects of the GWAS process could lead to biased test statistics, even under neutral conditions. We control for this possibility by sampling null SNPs so as to match the joint distribution of certain properties of the ascertained GWAS SNPs. Specifically, we were concerned that the minor allele frequency (MAF) in the ascertainment population, the imputation status of the allele in the HGDP datasets, and the background selection environment experienced at a given locus, as measured by B value \cite{McVicker:2009ic}, might influence the distribution of allele frequencies across populations in ways that we could not predict.

We partitioned SNPs into a three way contingency table, with 25 bins for MAF (i.e.\ a bin size of 0.02), 2 bins for imputation (either imputed or not), and 10 bins for B value (B values range from 0 to 1, and thus our bin size was 0.1). For each set of null genetic values, we sampled one null SNP from the same cell in the contingency table as each of the GWAS SNPs, and assigned this null SNP the effect size associated with the GWAS SNP it was sampled to match. While we do not assign effect sizes to sampled SNPs used to estimate the covariance matrix $\mathbf{F}$ (instead simply scaling $\mathbf{F}$ by a weighted sum of squared effect sizes, which is mathematically equivalent under our assumption that all SNPs have the same covariance matrix), we follow the same sampling procedure to ensure that $\mathbf{F}$ describes the expected covariance structure of the GWAS SNPs. 

For the skin pigmentation GWAS \cite{Beleza:2013cfa} we do not have a good proxy present in the HGDP population, as the Cape Verdeans are an admixed population. Cape Verdeans are admixed with $\sim 59.53\%$ African ancestry, and $41.47 \%$ European ancestry in the sample obtained by \cite{Beleza:2013cfa}  (Beleza, pers.\ comm.,\ April 8, 2013). As such, we estimated genome wide allele frequencies in Cape Verde by taking a weighted mean of the frequencies in the French and Yoruban populations of the HGDP, such that $p_{CV} = 0.5953 p_Y + 0.4147 p_F$. We then used these estimated frequencies to assign SNPs to frequency bins.

\cite{Beleza:2013cfa} also used an admixture mapping strategy to map the genetic basis of skin pigmentation. However, if they had only mapped these loci in an admixture mapping setting we would have to condition our null model on having strong enough allele frequency differentiation between Africans and Europeans at the functional loci for admixture mapping to have power \cite{Reich:2005iw}. The fact that \cite{Beleza:2013cfa} mapped these loci in a GWAS framework allows us to simply reproduce the strategy, and we ignore the results of the admixture mapping study (although we note that the loci and effect sizes estimated were similar). This highlights the need for a reasonably well defined ascertainment population for our approach, a point which we comment further on in the Discussion.

\section*{Acknowledgments}
We would like to thank 
Gideon Bradburd, 
Yaniv Brandvain, 
Luke Jostins, 
Chuck Langley, 
Joe Pickrell, 
Jonathan Pritchard, 
Peter Ralph,
Jeff Ross-Ibarra, 
Alisa Sedghifar, 
Michael Turelli
and 
Michael Whitlock
for helpful discussion and/or comments on earlier versions of the
manuscript. We thank Josh Schraiber and Otso Ovaskainen for useful
discussions via  \href{http://haldanessieve.org/2013/07/31/the-population-genetic-signature-of-polygenic-local-adaptation/}{Haldane's Sieve}.

\bibliographystyle{plos2009}
\bibliography{library,morelibrary}

\section*{Figure Legends}

\clearpage

\section*{Tables}

\begin{table}[!ht]
	\caption{The contribution of each geo-climatic variable to each of our four principal components, scaled such that the absolute value of the entries in each column sum to one (up to rounding error). We also show for each principal component the percent of the total variance across all eight variables that is explained by the PC.}\label{tab:eigenvector-table}
	{\begin{tabular}{c c c c c}
		\hline \hline
		Geo-Climatic Variable	&SUMPC1	&SUMPC2	&WINPC1	&WINPC2	\\
		[0.5ex]
		\hline
		Latitude	&-0.16	&-0.10	&-0.17	&-0.01 \\
		Longitude	&0.02	&0.12	&0.04	&0.05 \\
		Maximum Temp	&0.24	&-0.08	&0.17	&-0.03 \\
		Minimum Temp	&0.24	&0.07	&0.17	&0.08 \\
		Mean Temp	&0.25	&-0.03	&0.17	&0.03 \\
		Precipitation Rate	&-0.01	&0.16	&0.07	&0.32 \\
		Relative Humidity	&-0.06	&0.21	&-0.06	&0.34 \\
		Short Wave Radiation Flux	&-0.03	&-0.22	&0.15	&-0.13 \\
		Percent of Variance Explained	&38\%	&35\%	&58\%	&20\%	\\
	\end{tabular}}
\end{table}

\begin{table}[!ht]
	\caption{Climate Correlations and $Q_X$ statistics for all six phenotypes in the global analysis. We report $sign(\beta)r^2$, for the correlation statistics, such that they have an interpretation as the fraction of variance explained by the environmental variable, after removing that which is explained by the relatedness structure, with sign indicating the direction of the correlation. P-values are two--tailed for $r^2$ and upper tail for $Q_X$. Values for $\beta$ and $\rho$ are reported in Tables \ref{beta-table} and \ref{rho-table}.}\label{results-table}
	{\scalebox{0.8}{
	\begin{tabular}{c r r r r r r}
		\hline\hline
		Phenotype	&SUMPC1	&SUMPC2	&WINPC1	&WINPC2	&Latitude	&$Q_X$\\
		[0.5ex]
		\hline
		Height	&$-0.03\ (0.21)$	&$10^{-5}$ (0.99)	&$-0.008\ (0.52)$	&$\mathbf{0.086\ (0.035)}$	&0.009 (0.50)	&$\mathbf{86.9\ (0.002)}$\\
		Skin Pigmentation	&0.061 (0.073)	&0.003 (0.69)	&0.048 (0.13)	&$-0.008\ (0.51)$	&$\mathbf{-0.085\ (0.038)}$	&$\mathbf{79.1\ (0.006)}$ \\
		Body Mass Index	&$-0.034\ (0.19)$	&0.001 (0.82)	&$-0.022\ (0.31)$	&0.044 (0.14)	&0.031 (0.22)	&$67.2\ (0.087)$\\
		Type 2 Diabetes	&0.014 (0.40)	&0.012 (0.45)	&0.025 (0.27)	&$-0.006\ (0.573)$	&$-0.05\ (0.11)$	&39.3 (0.902)\\
		Crohn's Disease	&0.07 (0.062)	&$\mathbf{-0.099\ (0.022)}$	&0.0001 (0.94)	&$\mathbf{-0.09\ (0.039)}$	&0.01 (0.55)	&47.1 (0.68)\\
		Ulcerative Colitis	&0.03 (0.21)	&$\mathbf{-0.087\ (0.034)}$	&0.004 (0.67)	&$-0.049\ (0.12)$	&0.01 (0.43)	&48.58 (0.61)
	\end{tabular}
}}
\end{table}

\begin{table}[!ht]
	\caption{$Q_X$ statistics and their empirical p-values for each of our six traits in each of the seven geographic regions delimited by \cite{Rosenberg:2002ga}. The theoretical expected value of the statistic under neutrality for each region is equal to $M-1$, where $M$ is the number of populations in the region. We report the value of $M-1$ next to each region for reference.}\label{regional-table}
	{\scalebox{0.74}{
	\begin{tabular}{c r r r r r r r}
		\hline \hline
		&Europe (7)	&Middle East (3)	&Central Asia (8)	&East Asia (16)	&Americas (4)	&Oceania (1)	&Africa (6) \\
		[0.5ex]
		\hline
		Height &$\mathbf{32.6\ (<10^{-4})}$	&7.3 (0.07)	&$\mathbf{15.5\ (0.05)}$	&18.2 (0.33)	&4.2 (0.43)	&0.007 (0.94)	&5.4 (0.53) \\
		Skin Pigmentation	&9.7 (0.22) 	&$\mathbf{9.6\ (0.026)}$	&$\mathbf{23.4\ (0.002)}$	&13.8 (0.62)	&1.3 (0.89)	&0.38 (0.57)	&$\mathbf{16.2\ (0.011)}$\\
		Body Mass Index	&9.1 (0.24)	&1.6 (0.66)	&9.3 (0.32)	&$\mathbf{28.4\ (0.03)}$	&$\mathbf{13.1\ (0.016)}$	&1.2 (0.31)	&1.9 (0.94) \\
		Type 2 Diabetes	&2.0 (0.96)	&0.90 (0.83)	&8.1 (0.43)	&7.5 (0.96)	&8.0 (0.13)	&2.5 (0.15)	&2.5 (0.88) \\
		Crohn's Disease	&6.6 (0.47)	&0.87 (0.84)	&7.56 (0.48)	&15.5 (0.52)	&1.3 (0.88)	&2.5 (0.13)	&2.6 (0.82) \\
		Ulcerative Colitis	&8.4 (0.30)	&2.6 (0.48)	&10.9 (0.21)	&9.2 (0.907)	&0.43 (0.986)	&2.6 (0.12)	&3.5 (0.77)
	\end{tabular}
	}}
\end{table}

\newpage
\section*{Supplementary Figure Legends}

\setcounter{table}{0}
\renewcommand{\thetable}{S\arabic{table}}
\setcounter{figure}{0}
\renewcommand{\thefigure}{S\arabic{figure}}

\begin{figure}[!ht]
	\caption{Power of tests described in the main text to detect a signal of selection on the mapped genetic basis of skin pigmentation \cite{Beleza:2013cfa} as an increasing function of the strength of selection (A), and a decreasing function of the genetic correlation between skin pigmentation and the selected trait with the effect of selection held constant at $\delta = 0.13$ (B).}\label{skin-power-plot}
\end{figure}

\begin{figure}[!ht]
	\caption{Power of tests described in the main text to detect a signal of selection on the mapped genetic basis of BMI \cite{Speliotes2010} as an increasing function of the strength of selection (A), and a decreasing function of the genetic correlation between BMI and the selected trait with the effect of selection held constant at $\delta = 0.07$ (B).}\label{bmi-power-plot}
\end{figure}

\begin{figure}[!ht]
	\caption{Power of tests described in the main text to detect a signal of selection on the mapped genetic basis of T2D \cite{Morris:2012iu} as an increasing function of the strength of selection (A), and a decreasing function of the genetic correlation between height and the selected trait with the effect of selection held constant at $\delta = 0.08$ (B).}\label{t2d-power-plot}
\end{figure}

\begin{figure}[!ht]
	\caption{Power of tests described in the main text to detect a signal of selection on the mapped genetic basis of CD \cite{paper:2013iua} as an increasing function of the strength of selection (A), and a decreasing function of the genetic correlation between CD and the selected trait with the effect of selection held constant at $\delta = 0.05$ (B).}\label{cd-power-plot}
\end{figure}

\begin{figure}[!ht]
	\caption{Power of tests described in the main text to detect a signal of selection on the mapped genetic basis of UC \cite{paper:2013iua} as an increasing function of the strength of selection (A), and a decreasing function of the genetic correlation between UC and the selected trait with the effect of selection held constant at $\delta = 0.05$ (B).}\label{uc-power-plot}
\end{figure}

\begin{figure}[!ht]
	\caption{The two components of $Q_X$ for the skin pigmentation dataset, as described by the left and right terms in \eqref{Q_X_two_comps}. The null distribution of each component is shows as a histogram. The expected value is shown as a black bar, and the observed value as a red arrow.}\label{Q_X_components_skin}
\end{figure}

\begin{figure}[!ht]
	\caption{The two components of $Q_X$ for the BMI dataset, as described by the left and right terms in \eqref{Q_X_two_comps}. The null distribution of each component is shows as a histogram. The expected value is shown as a black bar, and the observed value as a red arrow.}
\end{figure}

\begin{figure}[!ht]
	\caption{The two components of $Q_X$ for the T2D dataset, as described by the left and right terms in \eqref{Q_X_two_comps}. The null distribution of each component is shows as a histogram. The expected value is shown as a black bar, and the observed value as a red arrow.}
\end{figure}

\begin{figure}[!ht]
	\caption{The two components of $Q_X$ for the CD dataset, as described by the left and right terms in \eqref{Q_X_two_comps}. The null distribution of each component is shows as a histogram. The expected value is shown as a black bar, and the observed value as a red arrow.}
\end{figure}

\begin{figure}[!ht]
	\caption{The two components of $Q_X$ for the UC dataset, as described by the left and right terms in \eqref{Q_X_two_comps}. The null distribution of each component is shows as a histogram. The expected value is shown as a black bar, and the observed value as a red arrow.}\label{Q_X_components_UC}
\end{figure}

\begin{figure}[!ht]
\caption{The genetic values for height in each HGDP population plotted against the measured sex averaged height taken from \cite{Gustafsson:2004hu}. Only the subset of populations with an appropriately close match in the named population in \cite{Gustafsson:2004hu}'s Appendix I are shown, values used are given in Supplementary table \ref{Gustafsson-height-table}} \label{Supp_fig_height_Gustafsson}
\end{figure}

\begin{figure}[!ht]
\caption{The genetic skin pigmentation score for a each HGDP population
  plotted against the HGDP populations values on the skin pigmentation
  index map of Biasutti 1959. Data obtained from Supplementary table
  of \cite{Lao:2007jf}. Note that Biasutti map is interpolated, and so
  values are  known to be imperfect.  
 Values used are given in Supplementary table
\ref{Supp_table_skin_pigmentation}} \label{Supp_fig_skin_pigmentation_Biasutti}
\end{figure}

\begin{figure}[!ht]
	\caption{The genetic skin pigmentation score for a each HGDP population
  plotted against the HGDP populations values from the 
  \cite{Jablonski:2000ds} mean skin reflectance (685nm) data (their Table 6). Only the subset of populations with an
  appropriately close match were used as in the Supplementary table
  of \cite{Lao:2007jf}. Values and populations used are given in Table
\ref{Supp_table_skin_pigmentation}} \label{Supp_fig_skin_pigmentation_JablonskiChaplin}
\end{figure}

\begin{figure}[!ht]
	\caption{The distribution of genetic height score across all 52 HGDP populations. Grey bars represent the $95\%$ confidence interval for the genetic height score of an individual randomly chosen from that population under Hardy-Weinberg assumptions}\label{ind-var-height}
\end{figure}

\begin{figure}[!ht]
		\caption{The distribution of genetic skin pigmentation score across all 52 HGDP populations. Grey bars represent the $95\%$ confidence interval for the genetic skin pigmentation score of an individual randomly chosen from that population under Hardy-Weinberg assumptions}\label{ind-var-skin}
\end{figure}

\begin{figure}[!ht]
	\caption{The distribution of genetic BMI score across all 52 HGDP populations. Grey bars represent the $95\%$ confidence interval for the genetic BMI score of an individual randomly chosen from that population under Hardy-Weinberg assumptions}\label{ind-var-BMI}
\end{figure}

\begin{figure}[!ht]
	\caption{The distribution of genetic T2D risk score across all 52 HGDP populations. Grey bars represent the $95\%$ confidence interval for the genetic T2D risk score of an individual randomly chosen from that population under Hardy-Weinberg assumptions}\label{ind-var-T2D}
\end{figure}

\begin{figure}[!ht]
	\caption{The distribution of genetic CD risk score across all 52 HGDP populations. Grey bars represent the $95\%$ confidence interval for the genetic CD risk score of an individual randomly chosen from that population under Hardy-Weinberg assumptions}\label{ind-var-CD}
\end{figure}

\clearpage
\begin{figure}[!ht]
	\caption{The distribution of genetic UC risk score across all 52 HGDP populations. Grey bars represent the $95\%$ confidence interval for the genetic UC risk score of an individual randomly chosen from that population under Hardy-Weinberg assumptions}\label{ind-var-UC}
\end{figure}

\clearpage

\section*{Supplementary Tables}
\begin{table}[!ht]
\centering
\caption{Genetic height scores as compared to true heights for populations with a suitably close match in the dataset of \cite{Gustafsson:2004hu}. See Figure \ref{Supp_fig_height_Gustafsson} for a plot of genetic height score against sex averaged height.}\label{Gustafsson-height-table}
\end{table}

\begin{table}[!ht]
\centering
\caption{Genetic skin pigmentation score as compared to values from Biasutti \cite{Parra:2004hk,Lao:2007jf} and \cite{Jablonski:2000ds}. We also calculate a genetic skin pigmentation score including previously reported associations at KITLG and OCA2 \cite{} for comparisson. See also Figures \ref{Supp_fig_skin_pigmentation_Biasutti} and \ref{Supp_fig_skin_pigmentation_JablonskiChaplin}.}\label{Supp_table_skin_pigmentation}
\end{table}

\begin{table}[!ht]
\centering
\caption{Conditional analysis at the regional level for the height dataset} \label{cond-height-region}
\end{table}

\begin{table}[!h]
\centering
\caption{Conditional analysis at the individual population level for the height dataset} \label{cond-height-ind}
\end{table}

\begin{table}[!ht]
\centering
\caption{Conditional analysis at the regional level for the skin pigmentation dataset} \label{cond-skin-region}
\end{table}

\begin{table}[!ht]
\centering
\caption{Conditional analysis at the individual population level for the skin pigmentation dataset} \label{cond-skin-ind}
\end{table}

\begin{table}[!ht]
\centering
\caption{Condtional analysis at the regional level for the BMI dataset} \label{cond-bmi-region}
\end{table}

\begin{table}[!ht]
\centering
\caption{Conditional analysis at the individual population level for the BMI dataset} \label{cond-bmi-ind}
\end{table}

\begin{table}[!ht]
\centering
\caption{Conditional analysis at the regional level for the T2D dataset.} 
\label{cond-t2d-region}
\end{table}

\begin{table}[!ht]
\centering
\caption{Conditional analysis at the individual population level for the T2D dataset.} 
\label{cond-t2d-ind}
\end{table}

\begin{table}[!ht]
\caption{Conditional analysis at the regional level for the CD dataset.}\label{cond-cd-region}
\centering
\end{table}

\begin{table}[!ht]
\centering
\caption{Conditional analysis at the individual population level for the CD dataset.} 
\label{cond-cd-ind}
\end{table}

\begin{table}[!ht]
\centering
\caption{Conditional analysis at the regional level for the UC dataset.} 
\label{cond-uc-region}
\end{table}

\begin{table}[!ht]
\centering
\caption{Conditional analysis at the individual population level for the UC dataset}
\label{cond-uc-ind}

\end{table}

\begin{table}[!ht]
	\caption{Corresponding $\beta$ statistics for all analyses presented in Table \ref{results-table}.}\label{beta-table}
	\end{table}

\begin{table}[!ht]
	\caption{Corresponding $\rho$ statistics for all analyses presented in Table \ref{results-table}.}\label{rho-table}
\end{table}

\end{document}